\documentclass[lettersize,journal]{IEEEtran}
\usepackage{amsmath,amsfonts}
\usepackage{algorithmic}
\usepackage{algorithm}
\usepackage{array}
\usepackage[caption=false,font=normalsize,labelfont=sf,textfont=sf]{subfig}
\usepackage{textcomp}
\usepackage{stfloats}
\usepackage{url}
\usepackage{verbatim}
\usepackage{cite}
\usepackage{color}
\usepackage{hyperref}
\usepackage{multirow}
\usepackage{makecell}
\usepackage{array}
\usepackage{booktabs}
\usepackage{graphicx}
\usepackage{epstopdf}
\usepackage{hyperref}
\usepackage[hyphenbreaks]{breakurl}
\graphicspath{{figure/}}

\begin{document}
\bstctlcite{BSTcontrol}
\title{{Large Language Models and Artificial Intelligence Generated Content Technologies Meet Communication Networks}}

\author{
Jie Guo,~\IEEEmembership{Senior Member,~IEEE,}
       Meiting Wang,\IEEEmembership{}
       Hang        Yin,\IEEEmembership{}
       Bin~Song,~\IEEEmembership{Senior Member,~IEEE,} \\
       Yuhao Chi,~\IEEEmembership{Senior Member,~IEEE,}
       Fei Richard Yu,~\IEEEmembership{Fellow,~IEEE,}
	  and  Chau Yuen,~\IEEEmembership{Fellow,~IEEE} 
\thanks{Jie Guo, Meiting Wang, Hang Yin, Bin Song, and Yuhao Chi are with the State Key Laboratory of Integrated Services Networks, Xidian University, Xi'an, 710071, China. }
\thanks{Fei Richard Yu is with the Department of Systems and Computer Engineering, Carleton University, Ottawa, K1S 5B6, Canada.}
\thanks{Chau Yuen is with the School of Electrical and Electronics Engineering, Nanyang Technological University, 487372, Singapore.}
}




\maketitle

\begin{abstract}
Artificial intelligence
generated content (AIGC) technologies, with a predominance of large language models (LLMs), have demonstrated remarkable performance improvements in various applications, which have attracted great interests from both academia and industry. 
Although some noteworthy advancements have been made in this area, a comprehensive exploration of the intricate relationship between AIGC and communication networks remains relatively limited. 
To address this issue, this paper conducts an exhaustive survey from dual standpoints: firstly, it scrutinizes the integration of LLMs and AIGC technologies within the domain of communication networks; secondly, it investigates how the communication networks can further bolster the capabilities of LLMs and AIGC. Additionally, this research explores the promising applications along with the challenges encountered during the incorporation of these AI technologies into communication networks. Through these detailed analyses, our work aims to deepen the understanding of how LLMs and AIGC can synergize with and enhance the development of advanced intelligent communication networks, contributing to a more profound comprehension of next-generation intelligent communication networks. 
\end{abstract}

\begin{IEEEkeywords}
LLMs, AIGC, communication networks, generative models, novel network architecture.
\end{IEEEkeywords}

\section{Introduction}

Large language models (LLMs) and artificial intelligence generated content (AIGC) technologies have recently advanced rapidly and are playing critical roles in a wide range of applications. For instance, ChatGPT, a groundbreaking example of LLMs, can generate valuable response content based on user input prompts and multiple conversations, providing a remarkable solution for multilingual machine translation, code debugging, text generation and recommendation systems, and so on \cite{OpenAI2023GPT4TR}. Its distinctive advantage lies in the powerful complex reasoning abilities, which facilitate harnessing extensive high-level knowledge derived from the backstage corpora.

As opposed to traditional human-generated content such as user generated content (UGC) and professional generated content (PGC) \cite{aigc}, 
AIGC technologies refers to a new automated paradigm in content generation, which uses generative artificial intelligence (GAI) models to produce multimedia digital content automatically. In state-of-the-art prompt-based GAI models, users can influence the output of the model by providing personalized descriptions. 
Currently, LLMs are emerging as one of the most important classes of GAI models. Compared to other small-scale GAI models like generative adversarial networks (GANs) and diffusion models, LLMs excel in two aspects. On the one hand, LLMs are powerful text generation models and the training process usually integrates with reinforcement learning from human feedback (RLHF)\cite{Ouyang2022TrainingLM} to further enhance the model understanding of user intent, enabling the realization of more authentic and realistic conversations. On the other hand, LLMs exhibit a larger parameter scale, capable of reaching several hundred to even over a thousand billion parameters. 
This substantial increase of model scale results in a significant enhancement in performance, commonly referred to as emergent capability \cite{wei2022emergent}, thereby yielding high-quality textual outputs and showing potential to serve as a generalist agent\cite{deng2023mind2web}.

\begin{table*}[htbp]\vspace{-0.8cm}
	\centering
        \renewcommand\arraystretch{1.5}
	\caption{Contribution comparisons between existing relevant surveys and our paper.}
	\begin{tabular}{{m{6em}<{\centering}m{5em}<{\centering}m{8em}m{38em}}}
		\hline\textbf{Research Field} &  \textbf{Year/Ref.} & \textbf{Topic Focus}&\textbf{Contributions} \\
		\hline
		\multirow{8}{6em}[-10ex]{\centering{GAI}}&2023/\cite{briefchatgpt} & ChatGPT&  {Overview of ChatGPT history and key technologies. Present the benefits and drawbacks of ChatGPT, as well as prospective challenges and developments in its application.} \\
  
		\cline{2-4}
		&2023/\cite{gaiaigc} &AIGC& {Survey on the evolution of generative AI models and recent advances including LLMs. Introduce relevant models and applications from the uni-modal and multi-modal perspectives.
}\\
		\cline{2-4}
		&2023/\cite{surveyLLM} & LLMs& Present a summary of existing LLMs by timeline, and summarize the characteristics of LLMs in four areas: pre-training, adaptation tuning, utilization, and evaluation. \\
		\cline{2-4}
		&2022/\cite{reasoning} &LLMs, reasoning capabilities& Survey on the reasoning capabilities of LLMs, outline the current state, relevant techniques, and assessment methods. \\
		\cline{2-4}
		&2023/\cite{eval} &LLMs, evaluation strategy& Survey on the evaluating LLMs and summarize what, where, and how they are evaluated. \\
		\cline{2-4}
		&2023/\cite{recomm} &LLMs, recommender systems& Summarize where and how to combine LLMs with recommender systems and present three key challenges. \\
		\cline{2-4}
  &2023/\cite{softwaretest} & LLMs, software testing& Summarize the distribution of testing tasks using LLMs, and analyze commonly used models, prompt types, and model inputs from the perspective of LLMs.\\
		\cline{2-4}
		&2023/\cite{surveyNLP} &LLMs, NLP& Summarize the availability, limitations, and challenges of using LLMs for various NLP downstream tasks and the influencing factors when deploying them. \\\cline{1-4}
\multirow{4}{6em}{\centering{Communication Networks}}
&2019/\cite{zappone2019wireless} &DL, physical layer&Survey on DL for wireless communication networks and the applications in operation and management of the physical layer.\\
\cline{2-4}
&2021/\cite{optim}&ML, end-to-end optimization&Survey on ML-based intelligent optimization method for end-to-end communication networks from data link layer to application layer\\
\cline{2-4}
&2021/\cite{edgeAI}&6G edge AI&Survey on scalable and trustworthy edge AI systems with integrated design of wireless communication strategies.\\
\cline{2-4}
\cline{1-4}
		\textbf{Our paper} & \textbf{2024}&  \textbf{GAI}, Communication Networks& \textbf{Overview of the intricate relationship between LLMs \& AIGC and communication networks. }Summarize the usages of LLMs \& AIGC technologies for communication networks, and the benefits of communication networks for LLMs \& AIGC. Present the applications, challenges, and future directions of LLMs \& AIGC in communication networks.\\
		\hline
	\end{tabular}%
	\label{survey}%
\end{table*}%

At present, LLMs and AIGC technologies have achieved significant advances in the fields of multi-modal data (text, image, video, etc.) production. For example, in text generation tasks, GPT-4 \cite{OpenAI2023GPT4TR} is capable of producing textual output utilized for dialogue systems, text summarization, and machine translation by processing input in the form of text or images. When tested with most professional and academic exams, the generated content performs at a human level and even ranks in the top $10\%$ on the simulated bar exam.
In image generation, the integration of the pre-trained text-image model with the diffusion model in DALL-E2 \cite{Ramesh2022HierarchicalTI} enables the real-time generation or editing of 1024$\times$1024 resolution images from natural language descriptions. 
Besides, progress has also been made in video generation. In the latest advancement, the Gen-2 model \cite{gen-2} can create approximately 3-second video clips based on textual and image prompts. 
Moreover, extensive explorations have been undertaken to apply LLMs and AIGC technologies in diverse application fields, including medical\cite{medical-1,medical-2}, finance\cite{finace-1,finace-2}, education\cite{edu-1,edu-2}, software engineering\cite{code-1,code-2}, and so on. 
Recently, noteworthy review papers\cite{briefchatgpt,gaiaigc,surveyLLM,reasoning,eval} have emerged detailing key technologies and developments in of AIGC and LLM. 
Other works delve into the pragmatic applications of LLMs in various fields, including natural language processing (NLP) downstream tasks \cite{surveyNLP}, recommender systems \cite{recomm}, and software testing tasks \cite{softwaretest}.

To date, the potential of LLMs and AIGC technologies in communication networks has not been thoroughly explored, which is a field worthy of research and development. Excitingly, it is in the hot phase of pre-researching key technologies for the 6th generation (6G) wireless communication networks, with the goal of a remarkable 10-fold reduction in latency and an impressive 10,000-fold increase in capacity compared to the current 5th generation (5G) wireless communication networks \cite{6g}. However, due to the diversity and complexity of supported service scenarios and data types in 6G communication networks, traditional communication technologies struggle to provide a perfect solution that can effectively address challenges such as high communication latency and missing transmission data. Existing surveys \cite{zappone2019wireless,optim,edgeAI} have shown the effectiveness and excellent power of machine learning (ML) and deep learning (DL) methods in communication networks. But existing solutions that combine traditional communication techniques with conventional DL algorithms suffer from issues such as poor semantic comprehension, limited performance, low universality, and the need for manual design.
It is worth noting that the powerful LLMs have the ability to fully utilize the massive amount of data in communication networks and have shown significant performance improvement, which is never achieved by traditional small-scale AI-empowered communication technologies. Besides, the outstanding generative fitting ability of
GAI models and the easily accessible AIGC data also outperform traditional communications in terms of fitting and data availability. 
Therefore, LLMs and AIGC technologies are promising to provide new insights and bring fundamental revolutions to the development of 6G communication networks. In summary, Table~\ref{survey} presents a clear comparison between the aforementioned existing literature and our paper.

Although LLMs are important components of AIGC technologies, they differ in many aspects. From the perspective of data modalities, LLMs are originally focused on processing textual data with excellent semantic understanding, whereas AIGC technologies deal with multiple modalities of data. From the perspective of processed tasks, LLMs can be regarded as a generalized interface that serves as a backbone model for multitasking, as they are well adapted to unseen data and tasks. In contrast, other AIGC models are more task-specific, providing superior solutions for existing communication issues based on excellent noise handling and generative fitting capabilities. Therefore, different LLM and AIGC techniques can be employed to improve the communication models according to different task requirements. Moreover, LLMs are converging with the AIGC technologies of various modalities to become multi-modal large models. 
Interestingly, advanced communication networks can be employed in turn to solve challenges associated with the development of LLMs and AIGC technologies, such as training and deployment. To make it clear, we provide the motivations of this paper as follows.

1) LLMs and AIGC technologies can be employed to enhance the communication systems from the following aspects.
\begin{itemize}	\item{\textit{Recovery of Missing Information.} 
In practical communication scenarios, the transmitted data is easily lost or corrupted due to interference and noise in the complex and varying wireless channels, as well as suboptimal coding and decoding strategies in the transceiver, making it difficult for the receiver to recover the information accurately. To overcome this challenge, LLMs and AIGC technologies can generate massive amounts of data \cite{Dai2023AugGPTLC} and simulate channel characteristics based on real-world channel environments\cite{condiGAN,Safety-Critical}, which enables them to effectively develop a robust coding and decoding strategy. The obtained channel information can be utilized for automatic network configuration. Meanwhile, by mining the potential semantic features of the received signals, the recovered information can be corrected by the tremendous generation capability of AIGC technologies and the semantic understanding capability of LLMs. On this basis, LLMs and AIGC contribute significantly to improving the robustness and accuracy of communication networks.
}

\item{\textit{Improvement of Multi-modal Task Performance.} 
Based on the excellent multi-modal semantic understanding of LLMs and AIGC, respectively, they can extract correct semantic information from various target tasks and massive data, thus improving semantic communication performance. 
Moreover, during semantic communication, the ability of the AIGC models to generate data from noise helps to mitigate the interference of channel noise on semantic features \cite{wu2023cddm}, and LLMs which serve as a knowledge base} can bridge the gap in local knowledge that exists between the transmitter and the receiver \cite{jiang2023large}.

\item{\textit{Privacy and Security Protection.} 
The numerous intelligent services in 6G communications pose challenges to privacy and security in communication. Utilizing LLMs and AIGC to conduct more sophisticated end-to-end encryption \cite{chen2022x} or optimize distributed learning architectures, such as federated learning \cite{chen2023federated} and splitting learning \cite{yao2022privacy}, are effective methods to prevent attackers from altering data or compromising user privacy.}
\end{itemize}

2) Advanced and widely deployed communication networks facilitate the practical application of AIGC technologies including LLMs, offering users higher quality services.
\begin{figure*}[!t]\vspace{-0.8cm}
	\centering
	\includegraphics[width=1\linewidth]{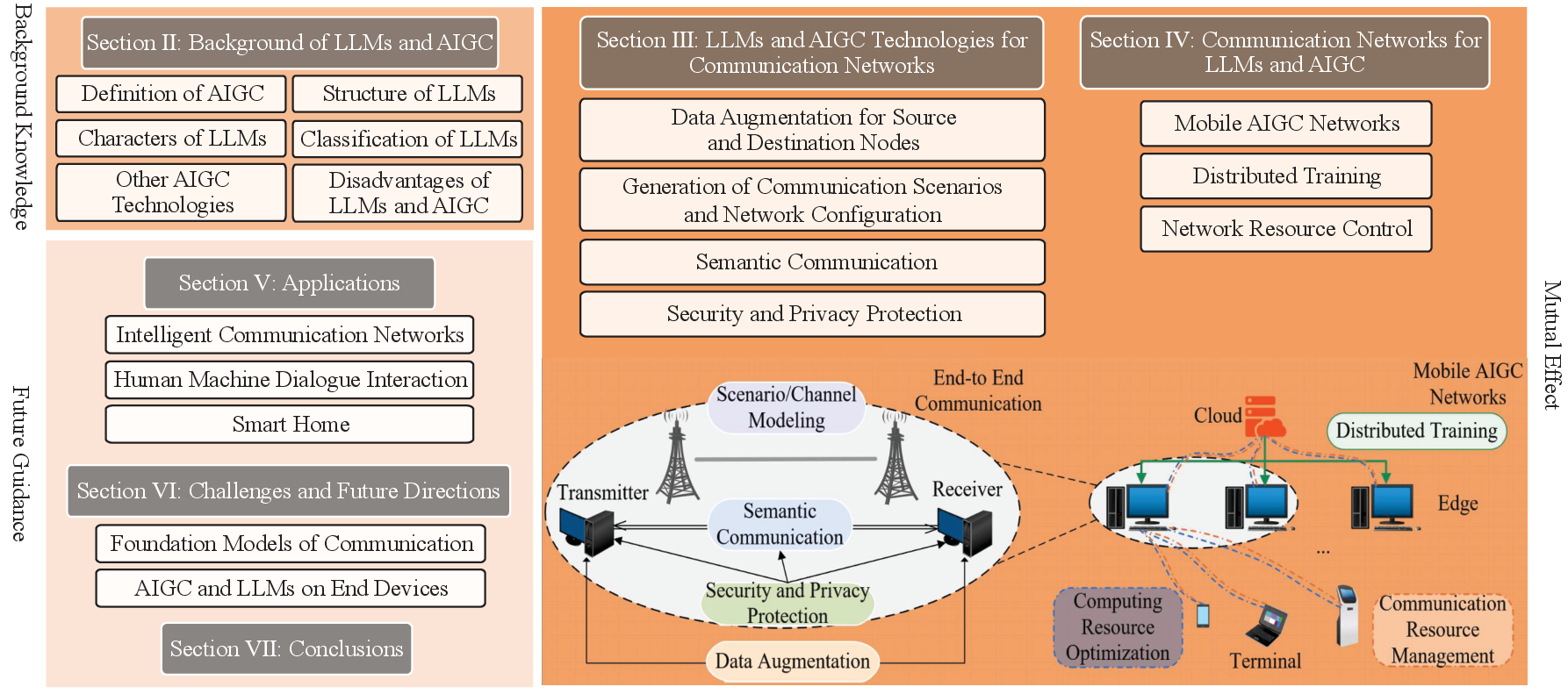}
	\caption{Organization structure of this paper.}
	\label{structure}
\end{figure*}

\begin{itemize}	
\item{\textit{Improvement of Training Efficiency.}
During the training phase of large-scale models, the training devices struggle to keep up with the rapid increase in model parameters. As a result, challenges in terms of training time and strategy arise. By implementing distributed training with wireless networks, training efficiency can be significantly improved by allowing multiple devices located in different locations to collaborate in parallel for large-scale model training \cite{megatron}}.
\item{\textit{Reduction of Latency.} 
LLMs and AIGC, due to their extreme complexity and huge scale, require massive computational resources for training or inference, which are difficult to execute on small terminal devices. Large-scale models are typically put on the cloud in order to deliver efficient AIGC services to users, and users access the cloud servers over the Internet to retrieve the services. However, this process encounters high latency and places a huge concurrent load on the cloud servers. A potential solution is applying the emerging cloud-edge-terminal collaboration in communication networks. The collaboration of devices with diverse capabilities and locations can rapidly alleviate the burden on device resources in AIGC services, effectively reducing data processing delays \cite{aigcservice}. Furthermore, with this architecture, AIGC models can be fine-tuned in real time depending on acquired data, resulting in better-focused services.}

\item{\textit{Support of Network Resources.} 
The control of network resources involving communication and computing resources can better allocate resources to the tasks of LLMs and AIGC \cite{kai2020collaborative}. Furthermore, the evaluation of LLMs and AIGC performance from the communication perspective helps to improve adaptation to communication tasks \cite{physical}.}
\end{itemize}

Based on the above analyses, we survey the use of LLMs and AIGC technologies in communication networks, summarize the benefits of using LLMs and AIGC techniques for communication networks, and optimize the deployment of LLMs and AIGC by communication techniques from the reverse perspective. Besides, some pertinent applications and challenges are presented. Our study provides readers with future directions and technological ideas for combining LLMs and AIGC with communication networks, which will contribute to the advancement of both AIGC and communication network technologies. The main contributions are summarized as follows: 

\begin{itemize}
	\item{We investigate how LLMs and AIGC technologies promote the advancement of communication networks. 
Specifically, the effects of LLMs and AIGC on content generation, performance enhancement, and security and privacy protection are explored.}
        \item{We investigate the benefits of communication networks for LLMs and AIGC technologies, such as cloud-edge-terminal collaborative network designs for application services, multi-device distributed training, and network resource optimization. These contribute to improved training and deployment of LLMs and AIGC.}
	\item{We summarize the applications related to the combination of communication networks and AIGC technologies, as well as the challenges and future directions.}
\end{itemize} 

The rest of the paper is organized as follows: Section \ref{s2} reviews the background knowledge of LLMs and AIGC. Section \ref{s4} introduces the innovations of applying LLMs and AIGC technologies to communication networks. Section \ref{s5} presents the support of communication networks for LLMs and AIGC. The applications are demonstrated in Section \ref{s6}. Challenges and future directions are presented in Section \ref{s7}, followed by the conclusions in Section \ref{s7}. Besides, the organization of the survey is shown in Fig.~\ref{structure}.

\vspace{-0.2cm}
\section{Background of LLMs and AIGC}\label{s2}
In this section, we review the definition of AIGC, the structure of LLMs, summarizing the characteristics, classifications, and the applications of LLMs and other AIGC models in the field of communications.
\vspace{-0.3cm}
\subsection{Definition of AIGC}
AIGC is usually used to refer to the content production method that uses GAI to automatically generate and create new content, including text, images, music, video, 3D interactive content, etc., based on training data and generative algorithmic models.
Existing AIGC technologies mostly use generative models such as GANs, variational auto-encoders (VAEs), flow-based models, and diffusion models, as well as the recently very attractive Transformer-based LLMs. 
As an efficient generative technology, AIGC is mostly task-oriented and integrates various AI technology approaches, including image processing \cite{Ramesh2021ZeroShotTG, Rombach2021HighResolutionIS, Ruskov2023GrimmIW} and sound processing \cite{Huang2023MakeAnAudioTG, Huang2023MakeAnAudio2T}, aiming at effective processing and collaborative generation of data in different modalities. The wide application of AIGC not only greatly improves the efficiency of content production but also changes the commercial mode of the multimedia industry.
\vspace{-0.8cm}
\subsection{Structure of LLMs}
Transformer \cite{Vaswani2017AttentionIA} model architecture is a stacked encoder-decoder structure primarily based on attention mechanisms. 
With the low computational complexity, parallel strategy, and multi-head self-attention mechanism, Transformer can handle input sequences and long-range dependencies faster and better.
Besides, Transformer has evolved into the backbone framework for numerous state-of-the-art language models (LMs), including GPT-3 \cite{NEURIPS2020_1457c0d6}, DALL-E2 \cite{Ramesh2021ZeroShotTG}, and Codex \cite{Chen2021EvaluatingLL}. 
When applied to the communication field, Transformer can be used to adaptively assign more weights and resources to more relevant historical channels for more accurate channel prediction\cite{transformer-channel} and explore the relevant features of the noisy signal and the reference signal in several potential feature domains by multi-attention to recover signals from noise\cite{transformer-signal}. In semantic communication systems, Transformer-based architectures have achieved remarkable success due to efficient extraction of semantic features from text and robustness to noise \cite{xie2021deep}.
\begin{figure*}[!t]
	\centering
	\includegraphics[width=0.8\linewidth]{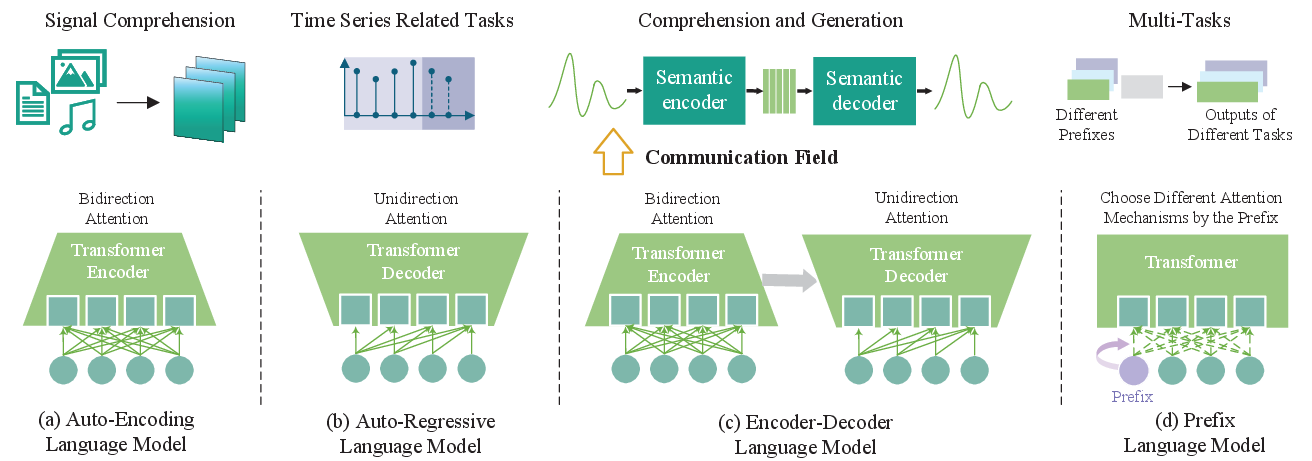}
	\caption{Schematics of the main structures in different LMs and
 corresponding applications in the communication field.}
	\label{LM}
\end{figure*}
LMs calculate the probability of linguistic sequences with the goal of predicting the likelihood of generating a sequence of words and, consequently, predicting the next (or missing) lexical element. The latest developments in LMs are based on the improvement and exploitation of the Transformer architecture, i.e., pre-trained language mode (PLM) and LLM. PLMs are large-scale LMs with millions of parameters that have been trained with large amounts of unsupervised data. Such models can be used as general foundation models and, after fine-tuning, customized for specific tasks. When the quantity of training data grows and the number of parameters reaches hundreds of millions, LMs evolve into LLMs. According to the model architectures and training methods, LMs can be categorized into the following four categories. In Fig.~\ref{LM}, we provide the schematics of the base structures that stack up to comprise different LMs.
\vspace{-0.2cm}
\begin{itemize}
    \item \textbf{Auto-Encoding Language Models.} 
Auto-encoding language models, such as BERT\cite{Devlin2019BERTPO} and RoBERT\cite{liu2019roberta}, predict masked tokens using a Transformer encoder with bidirectional attention, excelling in natural language understanding (NLU) tasks like contextual information extraction, which aids in feature extraction and text reconstruction.
However, this approach is unsuitable for generative tasks, as the masking mechanism limits learning correlations and creates inconsistencies between pre-training and fine-tuning. In communication systems, auto-encoding structures effectively leverage contextual information for tasks like signal reconstruction and compression\cite{auto-enco}.

\item\textbf{Auto-Regressive Language Models.} Auto-regressive language models use a Transformer decoder with masked self-attention\cite{Vaswani2017AttentionIA} to predict the next lexical element, excelling in natural language generation (NLG). The GPT series models\cite{Radford2018ImprovingLU,Radford2019LanguageMA} illustrate the effectiveness of this architecture.
However, they struggle with context utilization. Models like ELMo\cite{Peters2018DeepCW} try bidirectional interactions but still underperform in understanding tasks. 
This structure resembles auto regressive schemes in communication, which uses the past states to predict the next one, making it suitable for continuous communication processes like channel prediction tasks \cite{icc}.
\item \textbf{Encoder-Decoder Language Models.}
Encoder-decoder language models consist of two stacked Transformer blocks acting as encoder and decoder, similar to auto-encoding and auto-regressive models. This structure enables models like MASS\cite{Song2019MASSMS} and BART\cite{Lewis2019BARTDS} to perform both NLU and NLG tasks effectively, making them suitable for tasks requiring deep input understanding and high-quality output in communication, such as acting as codecs in semantic communication systems\cite{seq2seq}.
However, training these models requires large amounts of training data, resulting in few LMs based on the encoder-decoder architecture.

\item \textbf{Prefix Language Models.}
Prefix language models use the entire Transformer structure by splitting input into prefix and target text, encoding the prefix bidirectionally and the target sequentially. Models like UniLM\cite{Dong2019UnifiedLM} and ERNIE-M\cite{Ouyang2020ERNIEMEM} learn robust representations and generalize well with fewer parameters than encoder-decoder models in NLU and NLG tasks. However, they are less effective since the attention mechanism between encoder and decoder is vital. Prefix text enhances output controllability by describing tasks\cite{prefix-multitask}, enabling multi-task capabilities for text generation, summarization, and translation. Additionally, prefix conditioning supports multi-modal tasks, serving as unified supervision\cite{prefix-2,prefix-3}.
\end{itemize}
\vspace{-0.3cm}
\subsection{Characteristics of LLMs}
LLMs are LMs that contain billions (or more) of parameters, which can be adapted to a wide range of downstream tasks. 
As one of the AIGC technologies, by learning a large amount of text data and linguistic patterns, the LLM can generate realistic, coherent, and creative textual contents based on given context and prompt information, which is a powerful text generation tool. 

The characteristics of LLMs can be summarized in three specific points.

\begin{itemize}
    \item{\textbf{Large-Scale Parameters:} The number of parameters in LLMs is very large, reaching billions or even tens of billions, which allows the model to learn and represent complex linguistic patterns, including syntactic, semantic, and even contextual information, generating coherent and meaningful texts. 
    In addition, 
    the scale of training data is huge, coming from a variety of sources such as the Internet, books, and news. These pose an emergent phenomenon for LLM, namely, when the size of the model reaches a certain threshold, the model's performance and generalization ability appear to be significantly improved.}
    \item{\textbf{Multi-Tasking Capability:}
    LLMs are capable of handling various types of tasks, including text summarization, translation, sentiment analysis, etc. 
    Besides, there is an emergence phenomenon when the scale of LMs is increased. This phenomenon manifests as a substantial and abrupt improvement in model performance beyond a specified threshold. Therefore, LLMs have strong generalization abilities and exhibit performance far beyond that of typical small LMs when dealing with unseen tasks.}
    \item{\textbf{Two-Stage Training:}
    In the pre-training phase, LLMs are trained on large-scale general text datasets to learn the basic structure of the language and various common senses. Then, in the fine-tuning phase, the model is further trained on a smaller, more specific dataset, which is usually specific to a particular task or domain, such as medical text, legal text, or specific conversation data. 
    Fine-tuning allows the model to better understand and generate language for this specific domain and thus be better suited for the specific task.}
    \item{\textbf{Reasoning Capabilities:} 
    Chain of Thought (CoT) is an enhanced prompting technique that improves the performance of LLMs on complicated reasoning tasks, particularly mathematical problems. It uses prompts that outline specific output requirements for the model, enabling it to provide not only the final answer but also various intermediate reasoning steps or sub-goal outputs, which reveal the model’s entire inference process. This approach mimics human reasoning, whereby problems are solved step-by-step through a series of ordered thinking, analyzing, and reasoning steps. The use of CoT technique greatly enhances the reasoning ability of LLMs. Besides, it improves interpretability and controllability, allowing users to better understand the decision-making process of models.}
\end{itemize}\par
 \vspace{-0.2cm}
 Based on the above characteristics of LLMs, the application of LLMs in communication networks would bring numerous advantages. The massive number of parameters of LLMs, in conjunction with a large quantity of communication-related training data, makes it possible for LLMs to provide a comprehensive knowledge base in the field of communication network. In particular, firstly, a single {LLM-based model often competents for} multiple communication tasks. Based on abundant communication-related knowledge learned by LLMs, users can utilize various prompt engineering methods to make LLMs think step by step and then solve multiple specific communication-related tasks. 
Secondly, by low-cost fine-tuning {techniques}, excellent communication service quality can be achieved in {diverse scenarios without the consumption of computing and storage resources}. For example, low-rank adaptation of LLMs \cite{hu2021lora} efficiently fine-tunes the model in the form of a low-rank matrix by adding a small number of trainable parameters on the basis of the pre-trained LLM, effectively reducing the consumption of computing resources and storage resources.
Finally, the powerful heterogeneous data processing ability of LLMs enables them to better understand user’s requirements and network conditions, and the resources in communication networks, such as bandwidth, spectrum, and computing resources, can be uniformly scheduled and allocated by LLMs in accordance with different business requirements and network conditions.
\vspace{-0.3cm}
\subsection{Classification of LLMs}
Compared to the LMs with diverse structures, the existing LLMs have a more homogeneous structure, mostly based on the Transformer decoder structure. According to the development of the training method, LLMs can be divided into two types: base LLMs and instruction-tuned LLMs.

\begin{itemize}
\item \textbf{Base LLMs.}
Base LLMs have been trained to predict the next words (technically, to predict the next lexical elements) based on the given text input. Such models, for example, PaLM \cite{Chowdhery2022PaLMSL}, T5 \cite{Raffel2019ExploringTL}, and MT-NLG \cite{Smith2022UsingDA} are typically trained using huge volumes of untagged text from the Internet or other datasets to learn statistical patterns of language, but without a specific task orientation.

\item \textbf{Instruction-Tuned LLMs.}
Instruction-tuning is a training strategy for LLMs that tries to increase LLMs' capacity to complete new tasks by understanding task instructions, also known as prompts\cite{prompt}, without the use of explicit samples. 
The use of instruction tuning to fulfill the practical demands of users has been widely employed in existing LLMs, such as Bard5, GPT-4 \cite{OpenAI2023GPT4TR}, Panda LLM \cite{Jiao2023PandaLT}, InstructGPT \cite{Ouyang2022TrainingLM}, etc. Such models are often accompanied by manual annotations or instruction datasets. For example, the Flah-T5 \cite{Longpre2023TheFC} model uses the Flah-2022 \cite{Chung2022ScalingIL} instruction dataset to optimize the general language model T5 \cite{Raffel2019ExploringTL}, and the new model demonstrates superior capabilities. Moreover, prompt engineering can be used to design or automatically generate optimal instructions to better guide and fully utilize the model to obtain higher-quality results \cite{autoprompt}.
\end{itemize}
\begin{figure*}[!t] \vspace{-0.4cm}
	\centering
	\includegraphics[width=0.98\linewidth]{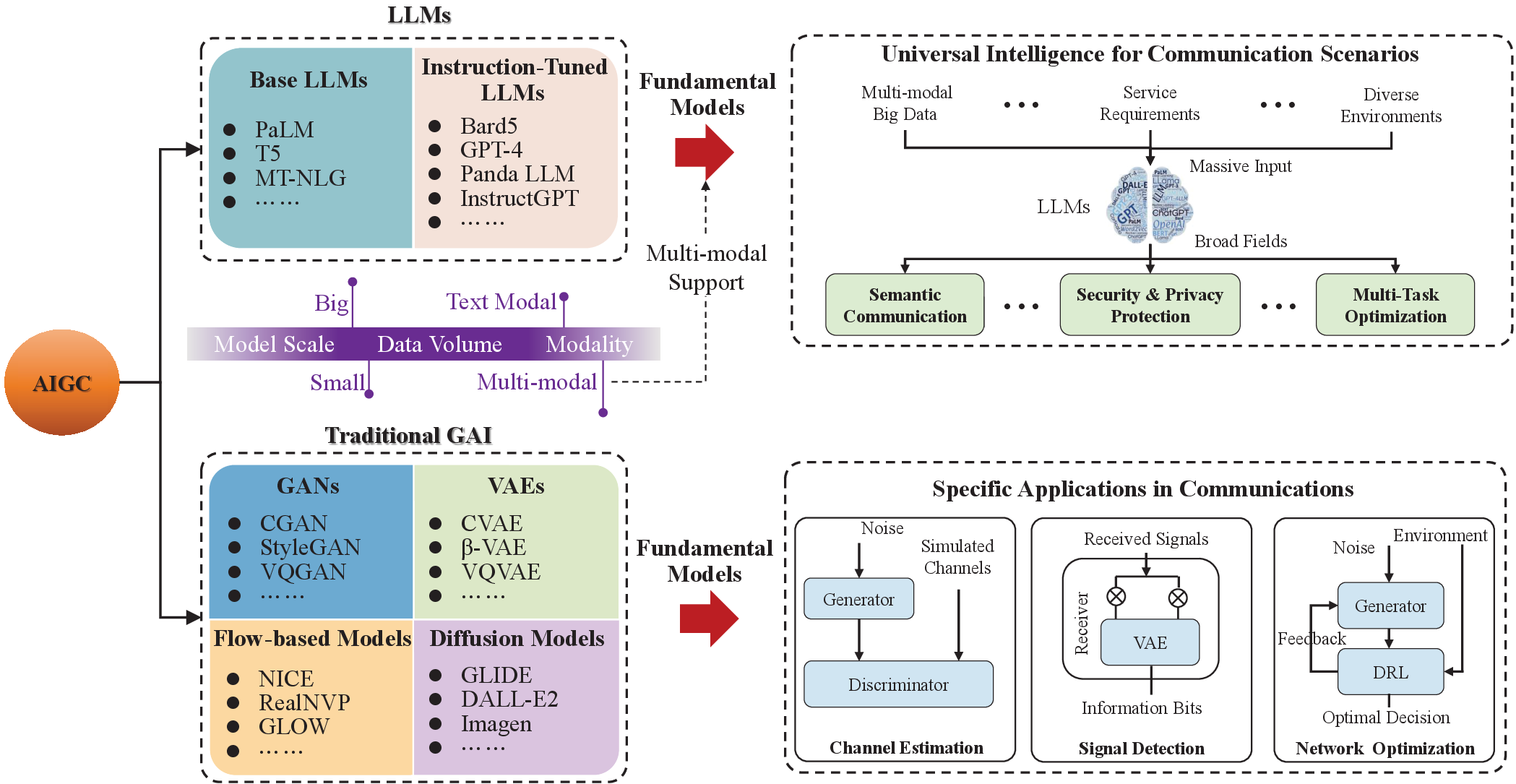}
	\caption{The classification and utilization of AIGC models in communication networks.}
	\label{section2}
\end{figure*}

\subsection{Other AIGC Technologies}
In addition to LLMs, AIGC models include other small-scale generation models involving text, images, audio, and so on. These models generate new content by learning the underlying patterns and structures of the data, making the generated content creative and in line with the needs of users. In this section, we introduce some generative models that are commonly used in AIGC. Besides, the classification of the AIGC models is shown in Fig.~\ref{section2}.

The common traditional GAI models are as follows.
\begin{itemize}
\item \textbf{GANs.} 
GANs\cite{Goodfellow2014GenerativeAN} consist of a generator and a discriminator that learn data distribution through adversarial training. Models like CGAN\cite{Mirza2014ConditionalGA} and StyleGAN v2\cite{Karras2019AnalyzingAI} have diverse applications in image, text, and audio. They generate clearer outputs quickly but struggle with discrete data and face training instability, limiting their use in communications. Some researchers enhance GAN training with additional information. In \cite{balevi2020high}, authors use simulated channel to generate spatial channel matrices.
\item \textbf{VAEs.} 
VAEs\cite{Kingma2013AutoEncodingVB,CVAE} are probabilistic models that consist of an inference network and a generation network, optimizing the variational lower bound to generate new data. They can extract latent information from signals in flat-fading channels, aiding signal integrity\cite{jaward2022joint}. While VAEs have potential for generating complex data, like images, their training is resource-intensive, and generated quality can vary due to sampling randomness.

\item \textbf{Flow-Based Models.} 
Flow-based models\cite{Dinh2014NICENI,Dinh2016DensityEU} use reversible transformations to connect original and generated data distributions. Although they allow easier access to latent features, they are computationally inefficient, taking much longer than GANs to generate images. They can benefit communication fields in tasks related to data encoding and modulation, as their ability to model complex data distributions enables more accurate signal representation and manipulation.

\item \textbf{Diffusion Models.}
Diffusion models, inspired by non-equilibrium thermodynamics, are latent variable models learned through a stationary process with high dimensionality. In\cite{du2023beyond}, their advantages over other generative models and the optimization guidelines are offered. Moreover,  in semantic communication systems, diffusion models can be employed for audio signal generation and information recovery \cite{grassucci2024diffusion}. 
\end{itemize}
\vspace{-0.6cm}
\subsection{Disadvantages of LLMs and AIGC}
At present, the application of LLMs and AIGC in communication networks is still not mature, and some shortcomings are still very challenging. The disadvantages of existing LLMs and AIGC are summarized as follows.
 \begin{itemize}
   \item\textbf{Limitations of Interpretability:} Due to the lack of interpretability of LLMs and AIGC, it is difficult for users to extrapolate in advance whether the currently generated solution can achieve the task objective efficiently and accurately.
    Moreover, the accurate analysis of the causes of inferior generation results is extremely difficult, which leads to users' distrust. To improve LLMs' interpretability, researchers employ knowledge graphs (KGs) to guide and analyze the generated content of LLMs at each step, thereby making the causal relationship at each step more explicit \cite{pan2024unifying}.
    \item \textbf{High-Cost of Computational and Storage Resources: }LLMs and AIGC need to be trained on a large number of GPUs for thousands or even tens of thousands of hours. Moreover, during inference, they also require floating-point computing capabilities ranging from trillions to tens of trillions or even higher per second. 
    In communication systems, this may put huge pressure on servers and networks, leading to performance degradation or service interruption. 
    To solve this problem, researchers are dedicatedly engaged in research in areas such as model compression, distributed computing, and hardware optimization. 
    \item \textbf{Inapplicability of Real-Time Applications:} 
    In some applications, autonomous driving for example, image data, radar perception data, sound data, etc. need to be processed in real-time. 
    However, the inference speed of recent LLMs ranges from about tens of milliseconds to a few seconds, which is obviously difficult to meet the real-time requirements. Moreover, the limited computing resources of end devices in communication systems further increase the response time of LLMs. 
\end{itemize}

\subsection{Lessons Learned}
LLMs, as one of the most representative GAI technologies in the AIGC field, provide potential abilities including generation, understanding, and creativity. 
Through syntactic and semantic analysis of texts, 
LLMs find extensive utility in NLP domains, spanning tasks like automated summarization, machine translation, and question answering, thereby bolstering the versatility of AIGC models across diverse scenarios.
Furthermore, LLMs significantly enhance the efficiency and quality in the AIGC field and can be combined with other GAI techniques to realize other modal tasks beyond language.
Moreover, the demand for AIGC services has actively contributed to the advancement of LLM technologies in practical application development. AIGC tasks can be used as evaluation benchmarks to assess the performance of LLMs. 
By comparing to real data and using feedback to improve the model, the quality of LLM-generated content can be improved. 
In communication networks, AIGC technologies can generate multi-modal data to improve specific communication tasks, while LLMs are expected to facilitate universal intelligence for various communication scenarios.

\section{LLMs and AIGC Technologies for Communication Networks}\label{s4}
In this section, we present the benefits of applying LLMs and AIGC technologies for communication networks from four aspects: data augmentation for source and destination nodes, channel and communication scenario generation, semantic communication, and security and privacy protection.

\subsection{Data Augmentation for Source and Destination Nodes}
For the source and destination nodes in communication networks, AIGC can be used for data augmentation to alleviate data insufficiency of DL models such as the joint source-channel coding model\cite{farsad2018deep}. On the one hand, since manual collection of data is highly expensive, the data used for training may be inadequate, making it difficult to improve model performance. On the other hand, communication data may be distorted or lost due to noise or interference during transmission, which results in difficulties in recovering the received information accurately. To solve such problems, data augmentation techniques can be used to enrich the existing datasets or to recover the damaged data based on the partial received data.

Compared with common DL models, the massive parameters of LLMs make them capable of storing a large amount of knowledge, and the pre-training based on a large-scale corpus gives them a powerful semantic perception ability that allows the use of rich prior knowledge to generate diverse data.
The form of data is known as its modality. For example, images, text, and audio are different modalities. Data within the same modality share a common structure, making the generation of single-modality data relatively straightforward. In contrast, generating data across multiple modalities is more complex, as it requires consideration of the semantic associations between data with different structures.
According to the modality of the generated data, augmentation techniques can be divided into two categories as follows.
\subsubsection{Unimodal Data Volume Expansion}
Generally, data augmentation considers unimodal data having a uniform data structure, which increases the number and diversity of training samples used for source and destination nodes.
There have been several studies using LLMs for textual data augmentation. For example, GPT3Mix\cite{Yoo2021GPT3MixLL} exploits the generative power of large-scale language models such as GPT-3 to blend existing samples to generate realistic-looking text samples.
 This method borrows ideas from MixUp\cite{zhang2017mixup}, a mixing-based augmentation technique that combines samples convexly in the image field and assigns the mixing process to a LLM.
AugGPT\cite{Dai2023AugGPTLC} uses the language understanding ability of ChatGPT to overcome the insufficient amount of domain-specific data by prompting ChatGPT to rephrase each input few-shot sentence into multiple similar sentences during the interaction process. 
These methods are simple attempts at data augmentation with LLMs, and the high-quality text generation shows excellent results on various text classification tasks.

In the field of image processing, using automatic enhancement or generative models for data augmentation is the current mainstream approach. In Autoaugment\cite{cubuk2018autoaugment}, traditional augmentation operations such as image cropping and panning are considered individual sub-strategies, and the combination of sub-strategies is searched by reinforcement learning to automatically learn the augmentation strategy from the data distribution. As the most representative deep generative model, GAN\cite{Goodfellow2014GenerativeAN} generates images from noise in an unsupervised manner by the generator, and the discriminator then adversarially tests the credibility of the generated results. The vast knowledge acquired by LLMs will benefit these small GAI models through knowledge transfer, thus improving the data augmentation quality in communication networks.\par

\subsubsection{Cross-Modal Data Supplementation}
Recently, several researchers have been studying cross-modal data augmentation methods that are not constrained to a rigid data structure. This means that data from noise-affected or lost modalities can be supplemented or recovered with modalities that have sufficient data volume in the communication nodes. Besides, the success of LLMs as language models for text generation is the cornerstone for the general use of multi-modal data generation. When combined with the AIGC models of the various modalities, the required data of different modalities can be generated leveraging LLMs given textual prompts.\par
In \cite{wang2022paired}, authors propose an augmentation framework for paired data to generate image-text data with the same semantics for cross-modal retrieval tasks. This framework constructs enhanced text by randomly replacing tokens in sentences, and uses the image generation model StyleGAN v2\cite{Karras2019AnalyzingAI} to construct enhanced images. 
Since the latent code of StyleGAN v2 is shown to have the capability to realize semantic disentanglement, the authors propose to project the text into this latent space and introduce a latent space alignment module to align the image latent codes with the corresponding text features. During generation, an augmented text is first constructed, delivered to the latent space alignment module to output the latent code, and finally provided to StyleGAN v2 to generate the augmented images. 
Also for image-text paired augmentation, the authors in \cite{hao2023mixgen} propose a concise hybrid generation method, MixGen. Given two existing image-text pairs, the new image is a linear interpolation of the two images by pixels, and the new text is obtained by concatenating the two texts. 
The use of hyperparameters for images blending is not easily generalized to different datasets, which can be combined with LLMs to optimize the mixing strategy.\par
Considering that the raw data may not be available or shareable, for example, the use of video data may have copyright requirements, a video-text paired augmentation method in feature space has been proposed in \cite{falcon2022textvideo}. Semantic classes with similar content are first defined based on the actions and objects extracted from the text, and then new videos and texts are created by blending two semantically similar samples belonging to the same class, where the blending is a linear interpolation of the embeddings of the samples.
This technique can be easily extended to different modalities because it works on latent representations. Moreover, this means that only pre-extracted features need to be shared and can be computed offline, which demands fewer resources. The ability of LLMs to grasp semantics can help feature-based cross-modal augmentation strategies improve semantic alignment. In this way, LLMs can assist in recovering damaged multi-modal data.\par

\begin{figure*}[!t]\vspace{-0.5cm}
	\centering
	\includegraphics[width=1\linewidth]{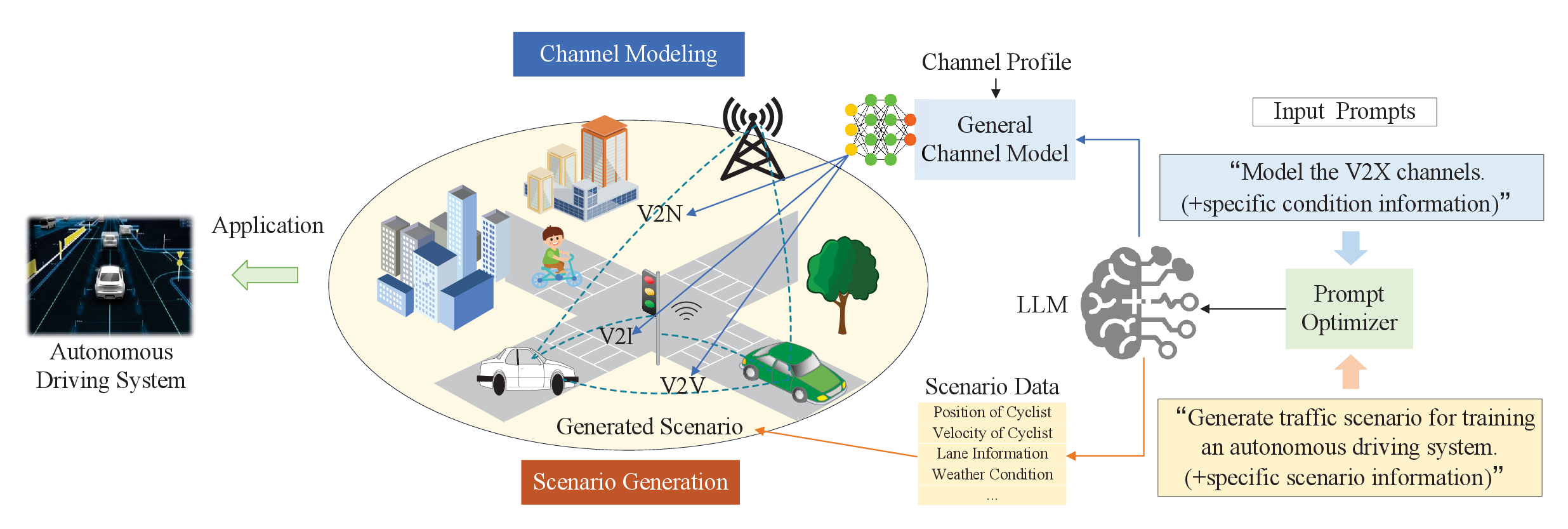}
	\vspace{-0.8cm}\caption{Channel modeling and scenario generation with LLMs (taking the traffic scenario as an example).}
	\label{generation}
\end{figure*}

\subsection{Generation of Communication Scenarios and Network Configuration}
Except for the generation of transmission data, DL can also be used to model and generate complex environments in communication, and even provide network configuration data automatically based on environmental information. Since the complete communication process is inevitably influenced by realistic environments, such as communication channels and driving scenarios, the influence of external environmental factors needs to be parameterized or simulated to optimize the communication quality in a targeted manner and improve the robustness of the communication system. Fortunately, the excellent ability of LLMs for the task of fitting and generation under complex conditions gives it great potential to deal with such problems.\par
To clearly demonstrate the proposed potential of LLMs in terms of channel and scenario generation, the combination of LLMs in the relevant techniques is shown in Fig.~\ref{generation}, using the traffic scenario as an example. In this figure, given textual input prompts for channel modeling and scenario generation separately, LLMs can model diverse channels and generate specific scenarios by leveraging their generative capabilities and extensive knowledge.

\subsubsection{Channel Modeling}
During the communication process, the signal sent by the transmitter will get attenuated by noise interference, fading, multi-path effects, and other effects in the channel medium before it is received by the receiver, resulting in distortion of the signal and the loss of information. 
In order to recover the affected signal and obtain complete and accurate information, it is necessary to model the channel with mathematical parameters to compensate for the channel effect at the receiver. However, classic statistical methods like the minimum mean square error algorithm or simple neural networks like ResNet \cite{resnet-channel-19} are too simplistic to model channels accurately in complex real-world environments.\par

In order to overcome the challenges mentioned above, a number of researchers have demonstrated the advantages of generative models as well as Transformer, the basis of language models, in channel modeling. When building an end-to-end communication system based on DL, the authors in \cite{condiGAN} propose that the problem of unknown channel functions impeding gradient backpropagation in DNNs can be avoided by simulating channel effects with the generative model CGAN. 
Taking the transmitted signal and the received pilot signal as conditions, the model can learn the output distribution of the channel and thus simulate complex channel effects. 

To predict the instantaneous channel state in fast time-varying environments, researchers often use RNN models, which suffer from error propagation due to their sequential nature. To address this issue, \cite{channel_T} proposes a parallel channel prediction model based on the Transformer, allowing multiple frames to be predicted simultaneously and overcoming error propagation. Additionally, the Transformer’s attention mechanism helps the model adaptively learn historical channel correlations to improve prediction accuracy. To further enhance channel estimation, \cite{channel_swinT} extracts temporal correlation features using 3D convolution and estimates channels with a CGAN, employing a Swin-Transformer as the generator's backbone for better feature extraction.

Transformers excel at learning temporal correlations, while GAI models demonstrate strong generative fitting capabilities, making them effective for channel modeling despite the complexities of high-speed changes. However, traditional DL models struggle with generalization when training data mismatches the application environment, necessitating specific training for different channels.
In contrast, large-scale models exhibit generalizability across tasks and strong few-shot learning capabilities. By using prompt as input, LLMs can describe various channel conditions, decoupling model parameters from specific contexts, which may enhance general channel modeling.

However, AIGC can introduce uncertainty, posing security risks in high-reliability tasks like channel modeling. To address this, prompt tuning can improve precision, ensuring prompts accurately convey requirements. Additionally, refining model structures or applying techniques like reinforcement learning can help prevent the generation of inaccurate information.

\subsubsection{Communication Scenario Generation}
Robustness and safety are key factors when evaluating the feasibility of deploying intelligent systems in the real world. 
In general, to improve the robustness of a DL model, adversarial attacks can be used to improve the sensitivity of the model to perturbations in the input. However, this method can only improve specific extreme conditions and can not make a comprehensive performance evaluation of the system.
In the scenario generation task, ordinary DL models are usually limited by the number of parameters and architecture to generate sufficiently diverse and realistic scenarios. Considering the ability of LLMs to understand complex semantics and the remarkable generation capability, they can make great progress in complex scenario generation.
In the following, we take the generation of autonomous driving scenarios as an example, because the reliability of automatic driving systems is related to the safety of people's property and even life.

The correct response of autonomous driving systems to safety-critical scenarios where accidents are highly likely to occur is essential \cite{cao2023dt}. However, such data is quite rare in the real world and thus scenario generation can product more diverse and safety-related key scenarios from existing scenarios.
Considering the scarcity and complexity of safety-critical scenarios, common DL models that estimate distributions based only on observed data are not able to generate such scenarios well. 
Recently, the authors in \cite{rR6} propose an LLM-driven scene generation framework that divides the driving scene generation process into three components: scene prompt engineering, LLM scene generation, and evaluation feedback adjustment. Furthermore, it enhances model performance through the use of CoT and experienced prompts. In addition, ChatSim \cite{rR7} utilizes LLMs as an agent collaboration framework, enabling users to perform highly flexible 3D driving scene editing through natural language instructions. The authors employ LLMs to leverage the vast external digital assets, thereby alleviating the demand for the task-specific scene data. Additionally, to effectively handle complex and abstract user commands, they utilize multiple LLM agents in collaboration, decoupling the overall scene simulation requirements into specific editing tasks and assigning these tasks to designated agents to provide realistic and detailed simulations.

Besides the topic of autonomous driving, the authors in \cite{rR8} discuss the challenges of using existing LLMs for 3D scene understanding. They have collected and constructed a set of instruction-response pairs specifically tailored for 3D scenes to address the scarcity of 3D scene-language pairs. Additionally, they propose an efficient prompt tuning paradigm that builds 3D multi-modal prompts to address the inefficient alignment between 3D scenes and language components. In \cite{rR9}, the authors investigate the current state of immersive imaging experiences and propose that the application of AIGC significantly enhances content personalization and dynamic adjustment capabilities, particularly in terms of visual effects and interactivity, highlighting its unique potential for optimizing immersive experiences.

\subsubsection{Automated Network Configuration}

Network configuration involves setting up and managing the parameters and protocols of network devices to ensure efficient, secure, and stable communication.
It is a dynamic process that requires continuous adjustment and optimization to adapt to changes in the network environment and requirements. Exploring the role of LLMs in network configuration is crucial for advancing automation, improving performance, and enhancing security in network management. \par
In \cite{need}, the authors find that using GPT-4 alone for generating router configurations results in significant topology, syntax, and semantic errors. However, with the proposed verified prompt programming method to correct these errors, effective configurations can be generated with a tenfold reduction in human input.
The authors in \cite{nconfig} have tested the effectiveness of LLM for automated network configuration, using LLM to automatically generate network configuration and routing algorithms based on specific requirements and policies described in natural language. In addition, they propose a set of design principles for LLM-based network configuration systems.
In order to optimize a vehicular communication system based on enhanced reconfigurable intelligent surface (RIS), the authors in \cite{ris} exploit the analytical capability of LLM to configure RIS in dynamically changing vehicular environments by combining crucial vehicular data, such as channel states, vehicle motion patterns, and quality of service requirements.
In \cite{IntentManagement}, the authors presents an architecture using LLMs for managing network services through natural language, which simplifies network configuration by translating user intents into actionable configurations. The approach covers the entire lifecycle of network intents, including decomposition, translation, negotiation, activation, and assurance. 
Real-world deployment demonstrates the practical applicability of this architecture, showcasing its ability to enable network configuration through natural language.


\subsection{Semantic Communication}
\begin{figure*}[!t]\vspace{-0.4cm}
\vspace{-0.2cm}
	\centering
	\includegraphics[width=0.98\linewidth]{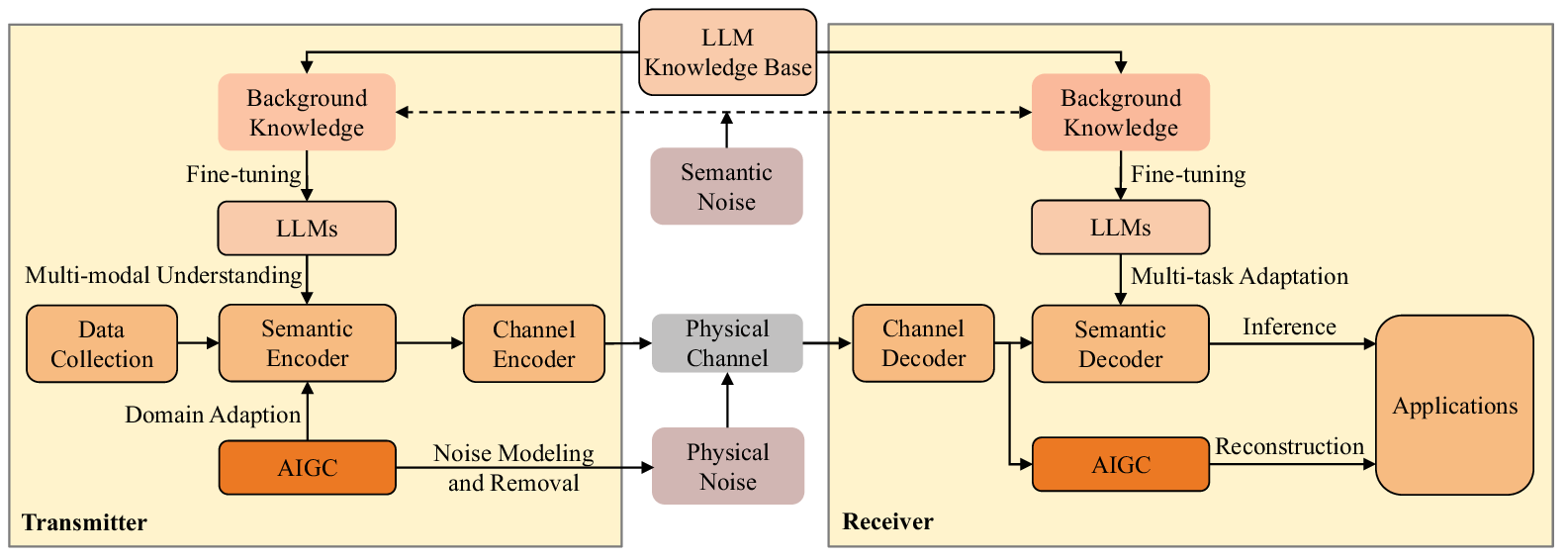}
	\vspace{0cm}\caption{A semantic communication system combining with LLMs and AIGC technologies.}
	\label{semantic}
\end{figure*}

Content-focused intelligent communication requires massive data transmission and quick system responses. Semantic communication is a core technology to address these challenges. 
The deployment of LLMs and AIGC can lead to a promising advancement in addressing the complex challenges like lack of background knowledge and high-quality semantic compression and reconstruction for semantic communication, facilitating more comprehensive and efficient information exchange, as shown in Fig.~\ref{semantic}.

\subsubsection{Semantic Transmission with Reduced Bandwidth}
Combining AIGC with semantic communication to transmit highly abstract semantic compression features can significantly lower the amount of physical bandwidth needed between the transmitter and receiver. This can help address the issue of bandwidth limitation in multimedia transmission channels.\par
Given the rapid development and excellent performance of LMs in NLP, researchers have first considered adopting LMs into communication systems to realize text-based semantic communication.
The authors in \cite{xie2021deep} preserve semantic information of sentences by first embedding sentences in a semantic space and then leveraging a Transformer-based encoder and decoder for text transmission. Compared with traditional communication, their proposed DeepSC achieves better performance in the low signal-to-noise ratio (SNR) condition. This demonstrates the effectiveness of using LMs for feature compression and recovery in communication processes.
In bandwidth-limited image transmission, the authors employ an approach where only the semantic segmentation map of the image is transmitted, and the receiver employs a pre-trained GAN network to reconstruct the realistic image \cite{lokumarambage2023wireless}. Compared with conventional image transmission, their approach achieves a compression ratio of approximately 20, which shows the superior performance of semantic level processing guaranteed by GAI. In \cite{li2021cross}, the authors introduce a cross-modal semantic compression framework that compresses images into compact textual feature representations at the transmitter and reconstructs the images by AttnGAN \cite{xu2018attngan} at the receiver. This framework based on GAI outperforms the widely used JPEG baseline with an exceptionally high compression ratio, significantly reducing the redundancy of transmitted data.

According to the multi-attention and probabilistic generation structure, the GAI models are easy to filter out the noise in the features, so as to realize accurate feature extraction and reconstruction. The authors in \cite{xu2023latent} create a denoising semantic communication scheme that utilizes adversarial learning, VAE, and diffusion models to reduce channel noise caused by channel characteristics at different SNRs during training and deployment stages. 
In this scheme, VAE serves as the encoder and decoder to extract semantic features and recover origin data. In the VAE latent space, the diffusion model functions as the semantic denoiser to model the channel noise and gradually removes the predicted noise in the process of backward denoising.\par
The authors in \cite{rR10} summarize the advantages of large AI models for semantic communication into three points, including accurate semantic extraction, rich prior knowledge, and robust semantic interpretation. They introduce a multi-modal semantic communication framework LAM-MSC based on large AI models, which utilizes multi-model LLMs to achieve multi-modal semantic alignment, and employs an LLM-based personalized knowledge base to mitigate semantic ambiguities during semantic extraction or recovery. Additionally, they estimate wireless channel state information using channel estimation based on CGAN to alleviate the impact of fading channels. In order to improve the transmission quality and the reliability of semantic reconstruction under low SNR conditions, LaMoSC \cite{rR11} designs a multi-modal semantic communication system that extracts text prompts using the LLM with extensive external knowledge, thereby overcoming the constraints of the knowledge base and limited generalization capability of traditional semantic communication systems. Additionally, since a significant quantity of external knowledge is incorporated into the encoding and decoding process, the correctness of reconstructed data is enhanced.
\begin{table*}[htbp]
\renewcommand\arraystretch{1.5}
  \centering
  \caption{Summary of references in semantic communication}
    \begin{tabular}{m{5em}m{3em}m{34em}m{16em}}
    \toprule
    \textbf{Topic} & \textbf{Ref} & \textbf{Main Idea} & \textbf{Revelations} \\
    \midrule
    \multirow{6}{5em}[-3.5ex]{Semantic Transmission with Reduced Bandwidth} & {\cite{xie2021deep}} & Construct a Transformer-based encoder and decoder for text transmission at the semantic level. & \multirow{3}{16em}[-1.5ex]{
    GAIs are effective for feature compression and recovery in communication processes since the semantic extraction and generation capabilities.} \\
          & {\cite{lokumarambage2023wireless}} & Compresse images into a semantic segmentation maps, which is recovered by GAN. &  \\
          & {\cite{li2021cross}} & Compress images into compact textual feature representations and reconstruct them using AttnGAN. &  \\
\cmidrule{2-4}          & \cite{rR10} & Use multi-modal LLMs for semantic alignment and CGAN for channel estimation to alleviate the impact of fading channels. & \multirow{2}{16em}[-0ex]{
GAI models have advantages in removing noise interference due to learning to generate features from noise.} \\
          & \cite{rR11} & Extract text prompts using the LLM to incorporate a significant quantity of outside knowledge into the encoding and decoding process. &  \\
\midrule    \multirow{5}{5em}[-6.5ex]{Adaptive Multi-Tasking Communication} & \cite{sanh2019distilbert} & Use the pre-trained DistilBERT model as a semantic recognizer to distinguish different users for one-to-many broadcast scenarios. & \multirow{3}{16em}[-2.5ex]{The generalization and semantic differentiation capabilities of LLMs facilitate multi-task and multi-user semantic communication.} \\
          & \cite{zhang2022unified}  & Develop a Transformer-based multi-task semantic communication by dynamically adjusting the length of transmitted features. &  \\
          & \cite{guo2023semantic} & Rank the semantic importance of input data by ChatGPT to optimize the transmission strategy for adapting to diverse requirements. &  \\
\cmidrule{2-4}          & \cite{zhang2022deep} & Transform transmitted data distribution into training data distribution by CycleGAN to improve compatibility with the original background knowledge. & \multirow{3}{15em}[-1.5ex]{
LLMs trained on large amounts of data can compensate for the different background knowledge required by different users and tasks.} \\
          & \cite{du2023generative} & Construct multi-modal prompts for inputs and reconstruct data from prompts with GAI. &  \\
          & \cite{jiang2023large} & Utilize a LLM as the knowledge base and personalized fine-tuning through user prompts to improve the accuracy of semantic feature extraction. &  \\
\bottomrule    \end{tabular}%
  \label{tab:addlabel}%
\end{table*}%
\subsubsection{Adaptive Multi-Tasking Communication}
Current DL-based semantic communication models are often only designed for specific modalities or tasks with poor generalization and universality, making them difficult to deploy in real application scenarios. To address this issue, semantic communication can utilize LLMs' strong adaptability to design a unified encoder-decoder model. In \cite{hu2022one}, the authors develop a Transformer-based semantic communication system for one-to-many broadcast scenarios. Specifically, the authors employ Transformer for semantic extraction and recovery and use pre-trained DistilBERT model \cite{sanh2019distilbert} for identification of different users based on the users' different content requirements, which is superior to the traditional communication and common DL-based semantic models in terms of bilingual evaluation understudy (BLEU) score.\par
In \cite{xie2022task}, a multi-task semantic communication system based on Transformer is investigated. The authors design a unified receiver to extract semantic features and three receivers to perform three specific tasks. In the cross-modal task, a new layer-wise Transformer is built for inter-modal information fusion by connecting each encoder and decoder layer. To some extent, this model is capable of handling diverse tasks; nonetheless, it still needs to be retrained for different tasks. To tackle this problem, the authors in \cite{zhang2022unified} develop a unified semantic communication system. In this system, a dynamic channel encoder is designed to adaptively adjust the transmission dimension according to different tasks and modalities. Meanwhile, at the receiver, the semantic decoder based on a unified Transformer takes the output features of the channel decoder and the task-specific query vectors as inputs, and the output features will be executed in subsequent layers. By utilizing the generalization ability of LLMs, a universal semantic decoder can be designed to facilitate multi-task and multi-user communication. 
In addition, the authors in \cite{guo2023semantic} introduce ChatGPT to rank the semantic importance of input data, which subsequently enabled the transmitter to allocate transmission power according to the importance ranking, thus implementing unequal error protection and optimizing the transmission strategy without retraining for different channels or tasks. This work indicates that ChatGPT can assist semantic encoders in adapting to diverse communication requirements.

Moreover, due to diverse local background knowledge in semantic communication, the issue of the semantic distortion between transmitter and receiver has emerged, i.e., semantic noise, resulting in a decline in the accuracy of information transmission. To handle semantic noise, the authors in \cite{peng2022robust} develop a calibration self-attention mechanism based on Transformer, which minimizes semantic ambiguity in text tokens. Besides, the authors employ adversarial training to overcome the impact of the adversarial semantic noise by maximizing the loss function to deceive the model during backpropagation. \par
In addition, model re-training is a way of reducing semantic noise generated by data heterogeneity in semantic communication, which nevertheless necessitates additional communication resources. In \cite{zhang2022deep}, the authors present a novel data adaptive network based on CycleGAN that improves compatibility with the original background knowledge for semantic understanding. This is accomplished by transforming transmitted data distribution into training data distribution without re-training.
In \cite{du2023generative}, the authors propose that the powerful generative capabilities of GAI allow semantic decoders to reconstruct source messages based on semantic prompts without the need to be jointly trained with semantic encoders. 
Therefore, they transmit only the multi-modal prompts for the face image transmission task. At the semantic decoder, a conditional diffusion model automatically generates the original image in accordance with the received prompts.
In \cite{jiang2023large}, the authors argue that large AI models help alleviate the problems of data heterogeneity and semantic ambiguity in multi-modal semantic communication. They constructed a multi-modal semantic communication framework based on large AI models, which realizes inter-modal semantic alignment and data transformation through multi-modal LLMs. To enhance semantic accuracy, a LLM is utilized as the knowledge base and personalized fine-tuning through user prompts to eliminate ambiguity in semantic extraction and reconstruction.

\subsection{Security and Privacy Protection} 

As 6G communication is closely associated with AI, a risk to user privacy and data security exists when large amounts of user data are collected and transmitted to train and update AI models. 
For example, insecure applications may send personal data to unknown AI systems, and attackers can steal user data by performing model inversion attacks through ML\cite{sun2020mach}. To address these problems, LLMs and AIGC models can be utilize for adaptive detection and control of potential threats in communication. 

\subsubsection{End-to-End Encryption and Secure Communication}
Traditional communication encryption approaches are insufficient to solve complicated scenarios in 6G communication, which calls for the employment of advanced LLMs and AIGC models to achieve end-to-end encryption and secure communication. 
The deployment of AI-enabled smart applications has 
significantly raised the stakes for sensitive data leakage, including but not limited to search history and bank accounts \cite{zhu2019deep}. 
To address this challenge, one potential solution is to combine end-to-end encryption with LLMs, which can analyze data and then encrypt it or obtain a unique key, ensuring data security while maintaining user privacy. In \cite{chen2022x}, the authors convert the Transformer model into a function comprised of homomorphic encryption operations via an approximation workflow.
This approach allows computation results to be decrypted only using the user's private key, ensuring user privacy when employing cloud services. In \cite{zhou2022textfusion}, the authors further utilize the Transformer with a dynamic token fusion mechanism for identifying confident token representations with lower uncertainty than a predefined threshold and recording their predictions. These token representations are then fused with adjacent token representations, while other token representations can be predicted on a server. In addition, the authors explore a deceptive approach to privatizing feature representations, which disrupts the one-to-one mapping between token representations and original textual data. As a result, the cloud server would only be able to obtain incomplete or perturbed feature representations, preventing it from reconstructing users' original data. 

During data transmission, the information received by the receiver may be intercepted or tampered with, leading to some security and privacy issues.
For instance, attackers may pose a risk of changing transmitted vectors to similar ones, which could cause output distortion at the receiver and be difficult to detect.
 To address this issue, the authors in \cite{lin2023blockchain} employ zero-knowledge proof techniques to record the transformation of semantic data, which can facilitate the sender in proving the truth of a statement to the receiver without revealing any relevant information. Furthermore, blockchain is adopted to monitor and validate the mutation of semantic data, thereby ensuring the authenticity and immutability of semantic vectors across the whole communication chain, from edge devices to virtual service providers. 
 To prevent attackers from obtaining the user's personal privacy during information transmission, the VQGAN-based method \cite{Yu2021VectorquantizedIM}, which uses autoregressive Transformers to synthesize high-quality images as mentioned in \cite{brkic2022privacy}, can be employed to remove identity information from images or videos. 
In\cite{luo2023encrypted}, to establish a universal anti-eavesdropping semantic encryption communication, the authors alternately train transmitters, receivers, and attackers using principles similar to GAN. The goal is to allow trained transmitters and receivers to beat good attackers. Meanwhile, this method also maintains the accuracy of the transmission. 

In summary, the user privacy protection scheme based on LLMs can be implemented by generating a unique key according form user and data information. Simultaneously, AIGC can encrypt the transmitted vectors at the transmitter end and then decrypt them at the receiver. These approaches can prevent the leakage of users' private information during data collection and model training processes.
\subsubsection{Distributed Learning-Based Secure Communication}

In the wake of increased privacy concerns among users, the traditional training methods based on centralized servers that require user data transmission pose significant privacy infringement, which motivates the development of federated learning (FL)\cite{mcmahan2017communication} as an alternative approach. FL enables edge devices to train local models using local data and then transmit only communication parameters to the remote servers for global model aggregation and updating. In addition, split learning (SL)\cite{gupta2018distributed} can be deployed to protect user privacy. 
In contrast to FL, SL shares the intermediate features of the data between devices rather than the data themselves, and allows edge devices and servers to train models separately without sharing models and parameters. To improve data protection, LLMs can be utilized to optimize FL frameworks and enable complicated data transformations.


There are some researches combining LLMs and FL to further protect user privacy. To address the conflict between the limited availability of public data and the requirement to protect users' private data, the authors in \cite{chen2023federated} propose a new structure called federated LLM, which includes three key components: federated LLM pre-training, federated LLM fine-tuning, and federated LLM prompt engineering. The federated LLM pre-training method involves the processing of source data by multiple clients and the design of LLM architectures through parameter selection and task design or fine-tuning of existing open-source LLM models.
The federated LLM fine-tuning integrates parameter-efficient methods into the FL framework, such as adapter tuning and prompt tuning. Federated LLM prompt engineering is conducted on sensitive private data, users share the parameters by uploading locally updated prompts, avoiding transmitting the original data. This structure addresses the critical issue of data privacy while maintaining model performance. \par

Moreover, in order to involve computationally restricted clients in the training of large models, the authors in \cite{tian2022fedbert} propose FedBERT, which follows an FL framework where pre-training tasks are distributed and executed in parallel among multiple clients. All clients and the cloud collaborate in training a pre-trained LLM by FL, with each client having the flexibility to fine-tune tasks. Each client only trains the head and embedding layers using its own data. The central server receives gradients from clients and updates the Transformer layers. When tested on RoBERTa \cite{liu2019roberta} and GPT-2 \cite{radford2018improving}, FedBERT effectively prevents the sharing of information between different clients while achieving superior performance. \par
To address the issue of privacy protection associated with LLMs in FL, the authors in \cite{zhao2023fedprompt} present a method that combines prompt tuning in a model split aggregation way to reduce the frequency of data exchange without undertaking retraining of the LLM, which prevents the leakage of raw data. In addition, their method has been proven to possess a high interception rate for backdoor attacks with poisoned training data.
 The authors in \cite{wang2023can} investigate the utilization of large-scale public data and LLMs to facilitate private training of FL across multiple devices. A public LLM is used to teach private on-device models through knowledge distillation, enhancing the privacy-utility trade-off for both large and small language models. Furthermore, the convergence speed of the model is improved by adopting public tokenizers from LLMs and sampling the public data based on its similarity to private data. Their approach effectively enhance private FL by leveraging public data and LLMs.

In the realm of secure SL, researchers have been exploring innovative approaches to ensure the protection of training data privacy \cite{erdougan2022unsplit}. The authors in \cite{yao2022privacy} propose a patch shuffling scheme, which leverages the robustness of the Transformer against shuffling and noise. 
To prevent input reconstruction, the position embedding layer is removed from the Transformer, followed by random batch shuffling and spectral shuffling of patch tokens. The randomness of patch tokens makes it harder for attackers to deduce input data from features.
Furthermore, the authors in \cite{xu2023shuffled} discover the permutation equivalence property of the Transformer in both the forward features and error gradients during backward propagation, which can be applied in SL. Consequently, edge devices can generate a random permutation order to function as the key encryption for raw data. Subsequently, the cloud leverages the permutation equivalence characteristic to enable training and testing using a one-time key. 
In order to protect user privacy and relieve the network burden, the authors in \cite{wang2023privacy} present a robust and efficient SL framework based on a masked pre-training Transformer. First, model parameters are initialized independently for both edge devices and cloud servers. After that, the edge device slices the image into patches, randomly selecting and masking a subset of patches using Fisher information, and encoding only the unmasked patches into the input encoder. 
 In conclusion, these advancements demonstrate a commitment to upholding data privacy in the SL framework, paving the way for more secure and efficient collaboration among edge devices and the cloud.

\subsection{Lessons Learned}
  The serialized inputs processed by LLMs exhibit a high degree of similarity to multimedia signals in the field of communication. Both tasks involve understanding and analyzing input information, followed by conditional generation based on these inputs. Therefore, applying LLMs to optimize communication networks shows great potential in areas such as data acquisition, information transmission, and privacy protection, which can enhance communication efficiency and improve content generation quality. 
  \begin{itemize}
    \item Content Generation:
        On the one hand, the information sent by the transmitter may be damaged during transmission in a bad communication environment. Missing information can be completed using LLMs and AIGC. Moreover, LLMs and AIGC can be used to generate real-world and reliable communication data. On the other hand, the generation of communication scenarios and network configuration with LLMs and AIGC can avoid wasting resources and time.
    \item Information Transmission:
        Information compression can reduce communication latency and communication consumption. Since LLMs trained on a large amount of data have a lot of background knowledge, they can enhance the semantic understanding of semantic encoders and semantic decoders. Moreover, communication systems using LLMs and AIGC can adapt to more complex and unstable communication scenarios.   
    \item Privacy Protection:
        In end-to-end communication, LLMs and AIGCs can protect user privacy by generating secret keys and encrypting transmitted information. Besides, combining LLMs with distributed learning schemes, such as FL and SL, can not only protect user privacy more effectively, but also improve the training efficiency.
\end{itemize} 

However, given the high requirements for information credibility in communication networks, it is essential to remain vigilant about the limitations of AIGC while leveraging its capabilities, especially concerning the issue of model hallucination. To enhance the reliability of AIGC, it is vital to improve the quantity and quality of data, optimize training strategies, or refine the model architecture.

\section{Communication Networks for LLMs and AIGC}\label{s5}
In this section, we present the details about how the communication network support the LLMs and AIGC in terms of mobile AIGC network, distributed training, and network resource control.

The mobile AIGC network is a collaborative framework consisting of various end devices which have AIGC cabibilities. The pervasiveness and flexibility of this network provides a solid physical foundation for subsequent distributed training, allowing data to be collected and processed over a wider geographic area, which in turn promotes model diversity and accuracy.
In the context of mobile AIGC networks, distributed training can fully utilize the computational resources in the network to achieve parallel processing of data and rapid iteration of models.
Besides, the limited nature of resources and the diversity of user demands in mobile AIGC networks constitute a pair of contradictions.
By dynamically adjusting network parameters, optimizing data transmission paths, and allocating computational resources, the network resource control strategy achieves fine management and efficient utilization of network resources. This leads to the reduction of energy consumption and hardware wear, thereby enabling mobile AIGC networks to better accommodate the continuously growing user and business requirements.
\subsection{Mobile AIGC Networks}

The efficient content generation capability of the AIGC models can bring about a significant increase in productivity for the content creation field. However, since the AIGC models are usually deployed on cloud servers, there is a high latency when a large number of users request remote access to the AIGC services. Using cloud-edge-terminal collaboration to deploy AIGC services, which constructs a mobile AIGC network, can make better use of the finite resources of computation, storage, and communication to achieve low-latency, personalized and privacy-secure AIGC services \cite{aigcservice}.
\subsubsection{Collaborative Architecture}
\begin{figure}[!t]\vspace{-0.2cm}
	\centering
	\includegraphics[width=0.98\linewidth]{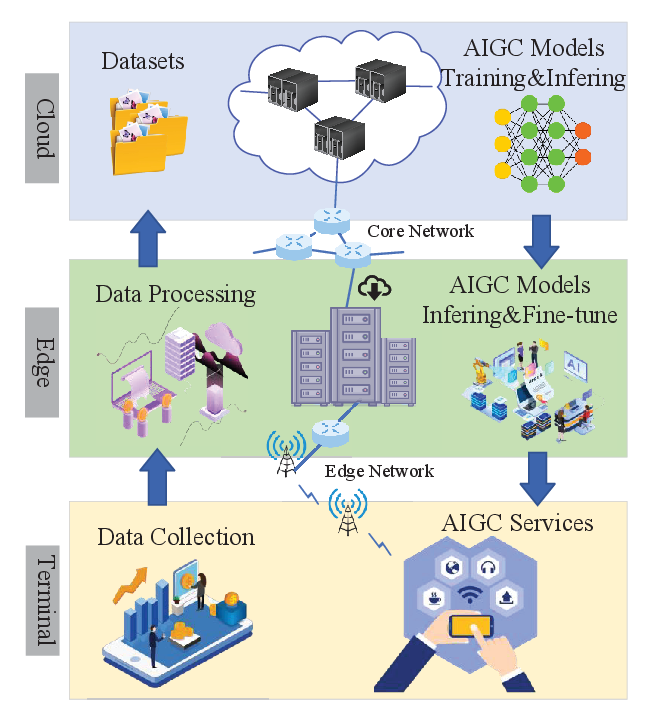}
	\caption{The cloud-edge-terminal collaborative architecture providing AIGC services.}\vspace{-0.2cm}
	\label{C_E_T}
\end{figure}

As shown in Fig. \ref{C_E_T}, the servers on the cloud side possess the most sufficient computational and storage resources in the cloud-edge-terminal collaborative architecture to execute model training that requires a large number of resources, which is the basis for providing AIGC services. Users can directly access the AIGC models in the cloud through the Internet. However, the service response latency is high due to the long distance. Besides, it is difficult for the cloud to deal with the massive accesses by a large number of users so frequently. According to the author's estimate in \cite{overview-E-C}, the current most popular AIGC service, ChatGPT, requires at least 350 GB of RAM and VRAM per day to run, costs up to \$600,000 per day in electricity if employing Nvidia A100 GPUs, and has considerable latency when relying solely on the computing resources available in the cloud. In order to alleviate the pressure on the cloud, the tasks with fewer resource requirements can be delegated to the edge side for execution. \par

Edge computing attempts to coordinate a large number of collaborative edge devices and servers to handle obtained data in close proximity, allowing AI models to benefit from aspects such as low latency, narrow bandwidth, and abundant data \cite{bg-edge}. 
{Based on different user requirements, different tasks would be realized through collaborating with edge devices and cloud services.}
Firstly, it is practical to deploy appropriate small models on the edge servers according to the large models trained in the cloud servers by utilizing migration learning, model pruning, and other methods to reduce the capacity of the models. Therefore, the AIGC services can be executed directly on the edge servers to reduce the response time and relieve the access pressure of the cloud considerably when the accuracy requirements are not too high. Moreover, based on the received user requests, the edge servers can fine-tune the model in real-time using the personalized data of users to generate content that is more in line with user needs and optimize the sense of user experience. In addition, as for user requests that cannot be handled by the edge side alone, the edge servers perform resource allocation and apply for resources from the cloud side.\par
On the terminal side, service requests submitted by users are forwarded to the edge. While the terminal devices are generally under-resourced to perform complex tasks, they can perform data collection and transmit the data to the edge server for data processing. 
Subsequently, the data is used at the edge for personalized tuning of the model or further transferred to the cloud for model training and updating.
When the resources permit, the terminal devices can perform model inference locally to satisfy privacy requirements and high latency requirements.
\subsubsection{Deployment of Models}
When deploying AIGC services based on cloud-edge-terminal collaboration, it is essential to strike a good balance between accuracy and resource consumption, leveraging the advantages of the AIGC models to achieve services that meet user needs.\par

In \cite{netgpt}, the authors propose the NetGPT architecture to deploy LLMs of different sizes at the edge and cloud for cloud-edge-terminal collaboration to provide users with personalized LLM-based generation services. For example, the open-source LLM LLaMa-7B \cite{llama} with 6.7 billion parameters is deployed in the cloud, and the GPT-2-base model \cite{Radford2019LanguageMA} with 100 million parameters is deployed at the edge and fine-tuned to be used for the prompt expansion task, in which expand the concise textual prompts submitted by the user to generate a composite prompt containing the user's location information on the basis of the location information obtained from the communication messages. Subsequently, the original and composite prompts are delivered to the cloud for inference to generate a personalized answer based on the user's location.\par
Given that the resources of edge devices may not be sufficient to allow parallel loading of various large-scale pre-trained models with diverse capabilities and performing inference, the authors in \cite{r9} present a joint model caching and inference framework. According to user demands, the edge server in this architecture allocates memory and compute resources flexibly and decides whether to ask for job offloading to the cloud. Consequently, the service cost, latency, and accuracy loss are diminished while fulfilling the user's requests for generative AI services within the resources available at the edge.
In \cite{UVA}, the authors investigate a framework for AIGC services with UAVs as edge servers. To reduce latency, the UAVs are outfitted with a pre-trained AIGC model and then fine-tuned using user-uploaded data. After receiving requests from users, AIGC reasoning is performed according to the prompts to provide low-latency and highly reliable AIGC services. The flexibility of the UAV greatly improves the scope of AIGC service delivery.

\subsection{Distributed Training}
\begin{figure*}[!t]
	\centering
	\includegraphics[width=0.95\linewidth]{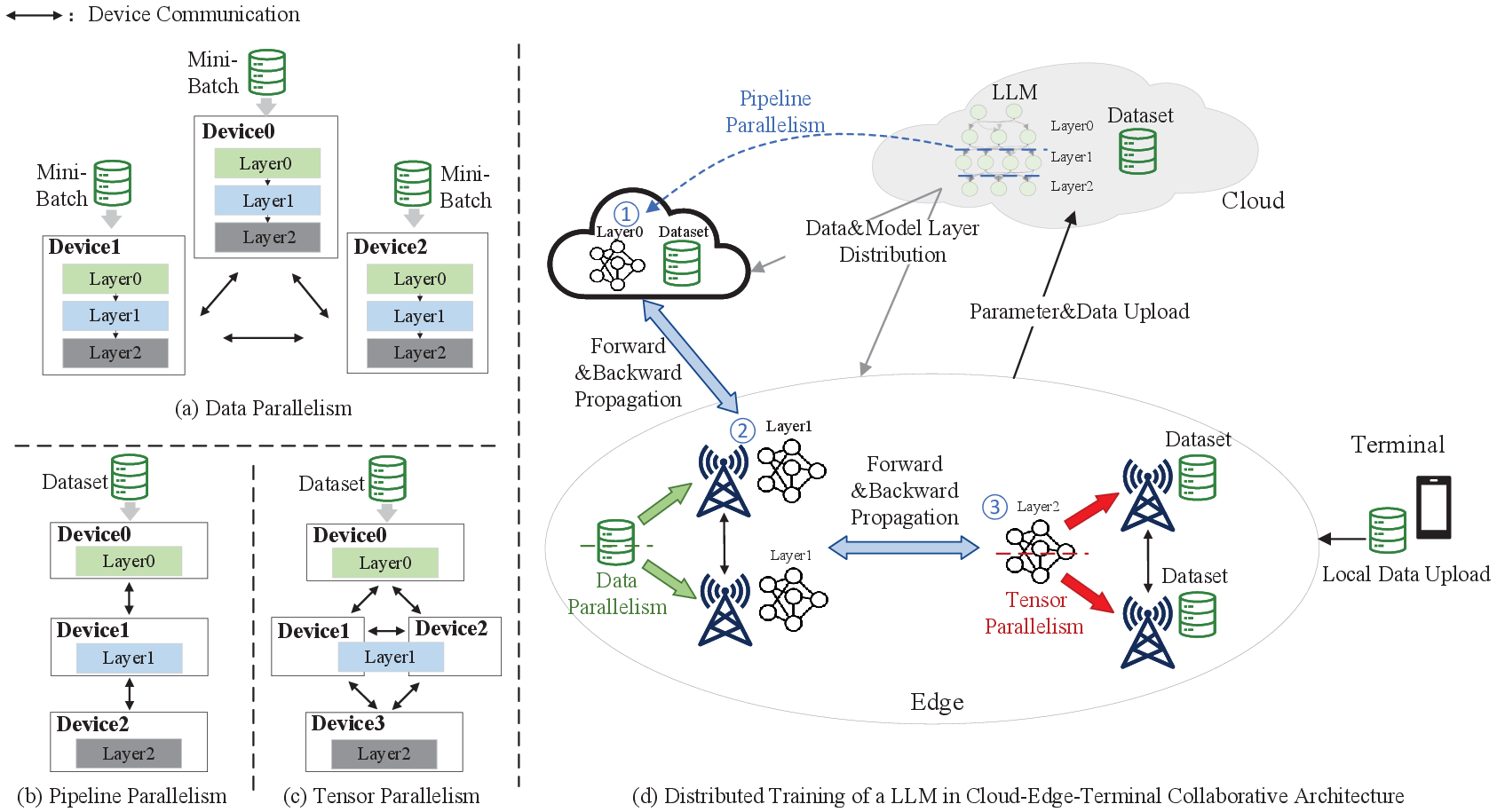}
	\caption{An illustration of three parallelism strategies of distributed training and the utilization for a LLM.}
	\label{distributed}
\end{figure*}
The large number of parameters within LLMs results in their substantial demand for computing resources during the training process. To solve this problems, 
distributed training emerges as an effective approach to alleviate this strain by partitioning and distributing training tasks across multiple computing devices, facilitating collaborative training through device communication. 
\subsubsection{Parallelization Strategies}
The distributed training approach can be categorized into two types: data parallelism and model parallelism. Moreover, model parallelism can be further divided into pipeline parallelism and tensor parallelism. The distinctions among these three parallel methods and the utilization example for LLM training are illustrated in Fig.~\ref{distributed}.

Data Parallelism partitions the training data into multiple mini-batches, concurrently processes distinct mini-batches on separate computing devices, and updates model parameters via gradient aggregation. In this paradigm, each computing device necessitates a complete model copy, and communication is primarily engaged during backward propagation for model gradient transfers. \par
Model Parallelism, on the other hand, segments the model and assigns computational tasks for discrete sub-models to different computing devices. Depending on the manner of partitioning, this technique bifurcates into tensor parallelism and pipeline parallelism. Pipeline parallelism refers to inter-layer model parallelism, wherein the model is fragmented by layer, and distinct layers' computational tasks are conducted on disparate computing devices. Upon completion of its computation stage, each device transmits intermediate results to the subsequent computing device for further processing, resulting in diminished communication overhead yet introducing additional idle waiting periods. \par
Tensor parallelism, in contrast, represents intra-layer model parallelism, where tensors within network layers are divided into chunks and computed independently on separate devices. Communication takes place during both forward and backward propagation at each layer, leading to frequent and relatively substantial data exchanges. 
For hyper-scale language models, a combination of multiple parallel strategies is typically employed to achieve more efficient training. This necessitates a more flexible architecture for model training design and proficient communication mechanisms.
\subsubsection{Training Implementation}
For large-scale models based on Transformer architecture, Megatron-LM \cite{megatron} proposes a simple and efficient model parallel training method. Exploiting the characteristics of the Transformer structure, the intermediate tensors in the multi-layer perceptron layer and the self-attention layer are split respectively. It is actually to utilize the principle of matrix multiplication and chunking to design the chunking method that does not change the computation result. Also, it can be combined with data parallelism and pipeline parallelism by distributing the layers to different devices.
In the model parallel approach GPipe \cite{GPipe}, the model is partitioned into different layers across computing devices and executed in a pipelined manner. In order to improve the device utilization, the mini-batches of data are further split into micro-batches, thus reducing the waiting time during sequential execution.\par
In \cite{Optimus}, the authors contend that the intra-layer partition approach, similar to that of Megatron-LM, is easy to implement but faces challenges such as memory redundancy and suboptimal communication efficiency. To address these issues, an efficient and scalable framework named Optimus is introduced. Within Optimus, two essential enhancements are achieved. Firstly, a highly effective tensor parallelism approach is implemented using a two-dimensional matrix partition technique, which effectively alleviate the memory bottleneck. Secondly, buffers are pre-allocated for communication data and transmitted parameters, thereby optimizing memory management for improved performance.\par
To achieve better training results, 3D parallel frameworks that integrate data parallelism, pipelined model parallelism, and tensor model parallelism are employed. However, 3D parallel frameworks recently suffer from two problems: the requirement of manual modification for training frameworks and the lack of resource utilization. To ameliorate these problems, the authors in \cite{Merak} propose the Merak framework, which provides an automatic model partitioner and an API to automatically deploy 3D model parallelism and assist developers in modifying the training framework. Besides, a 3D parallel runtime engine is designed to improve the computational resource utilization and communication concurrency.
In \cite{autopipe}, considering the impact of model partition balance on pipeline efficiency and the inevitable waiting time overhead during pipeline startup, the authors propose a fine-grained partitioner and a micro-batch slicer. Thereby, a balanced pipeline partitioning scheme is automatically and quickly generated, and micro-batches are uniformly split to reduce pipeline startup overhead, leading to a significant improvement in pipeline efficiency.
\vspace{-0.3cm}
\subsection{Network Resource Control} 
With the rapid development of DL and the growth of data, the demand for network resources has also increased significantly \cite{cao2024softwarized,cao2023resource}, particularly for the high resource consumption AIGC models, making effective management and allocation of network resources crucial. 
Specifically, we first study two aspects: communication resources and computational resources, and provide a comprehensive analysis and discussion of their respective impacts and strategies. 
Evaluation metrics aid in measuring the operational status and performance of the network so that the network resource control strategy can be optimized based on their changes.

\subsubsection{Communication Resources Management}
The transmission of AIGC data may require greater bandwidth, especially when processing large-scale media data. To improve resource usage, communication resource management must consider the transmission requirements and perform proper allocation schemes.

In edge networks with limited communication resources, Kai \emph{et al.} \cite{kai2020collaborative} proposed a scheme that comprehensively considers transmit power allocation, trasmission rate, computation offloading strategy, and computational resources. By addressing the non-convex optimization problem to minimize total latency, this approach enhances the processing efficiency of collaborative tasks between the cloud and edge. Such approaches are especially important for resource-intensive AIGC services since they balance limited transmission capacity with large amounts of service data, which improve the quality of AIGC task delivery.\par

The rise of AIGC models gives users more options, but sending requests to multiple service providers can waste resources. A common yet ineffective approach is for users to send tasks to the AIGC service provider (ASP) believed to produce the best content. However, limited bandwidth and task interruptions can overload large AIGC models, and users often need to make multiple requests to assess content quality, further straining network resources.
To address this issue, Du \emph{et al.}\cite{du2023enabling} explore optimal ASP selection at the edge. They utilize a participant policy network with softmax activation to determine the selection probabilities for each ASP under resource constraints. A deep reinforcement learning framework then chooses an ASP based on these probabilities, rewarding high-quality content and penalizing congested models. This approach allows for more efficient bandwidth allocation and improves management of communication resources.

\subsubsection{Computing Resources Optimization and Computation Offloading} 
The AIGC models especially LLMs are complex and large-scale, and their training process usually involves a large number of parameters and large-scale datasets. 
Therefore, efforts are being made to optimize limited computational resources and alleviate the computational pressure on the cloud and edge to support the model training and inference.

Offloading lightweight tasks such as model inference to edge devices can alleviate the burden on cloud computing resources. 
Collaboration between edge devices can give full play to the potential of each device and improve the performance and efficiency of the entire system.
In\cite{du2023exploring}, the authors propose a distributed AIGC framework based on collaboration, which utilizes the collaboration between edge devices to optimize computing resources. In the training phase, a large amount of data is used to train the AIGC model to generate high-quality content. Then the edge server collects user requests and calculates the required resources to ensure efficient resource utilization. The system also identifies differences and similarities in user requirements, groups users with similar task requirements, and customizes shared denoising steps for each group. In the inference stage, each user group with similar task requirements performs a shared denoising step, and the denoised output is passed to their respective devices to complete their tasks independently. In this way, we can not only save resources but also maintain user privacy, while the user personalization will be more flexible.

Due to the diverse application scenarios and requirements, users would send requests to AIGC services in various forms, such as text and image. In certain scenarios, conveying information like gestures through language can be challenging, which may affect the quality of content generated by AIGC models. 
Based on the integration of wireless perception (WP) and AIGC, the authors in \cite{wang2023guiding} present a unified WP-AIGC framework to improve the quality of content generation. By utilizing WP information as input to the AIGC model, the corresponding digital content is generated. However, both WP and AIGC require computing resources, and how to allocate resources reasonably needs to be solved. 
According to the total variation and blind/reference image spatial quality estimator, they further monitor the quality of generated images and feed them back to WP-AIGC. WP-AIGC will adjust available resources after receiving feedback to achieve a balance between WP and AIGC resource usage.

\subsubsection{Evaluation Metrics}
Consideration of appropriate metrics can aid in the optimization of mobile AIGC networks and the provision of better AIGC services. And some of the traditional metrics can be considered in a new way from the standpoint of the AIGC. In the following, we introduce the relevant metrics from two aspects: the quality of AIGC service and communication transmission. Moreover, these two types of metrics can mutually promote each other to some extent. On the one hand, the AIGC service metrics call for an improvement in the quality of the generated data, which can be used to recover damaged data during communication. Also, higher-quality generation means that the content may be more compact and abstract, reducing the amount of data to be transmitted and thus easing the pressure on communication transmissions.
Moreover, reduced distortion in communication transmissions contributes to the accurate transmission of AIGC service data, and lower latency and high bandwidth utilization support the rapid response of the AIGC service. 

\emph{a) Metrics Related to AIGC Services Evaluation:} There are some metrics usually used in specific communication tasks, which benefit AIGC services. Based on user demands for service accuracy, resource and task allocation can be optimized by integrating these metrics with considerations of network bandwidth and resource availability, to strike a balance among multiple demands.\par
\begin{itemize}
\item \textbf{BLEU}. BLEU \cite{bleu} is a metric for evaluating the quality of machine translation and text generation tasks.
This metric calculates the {difference} between machine-generated texts and professional human-written reference texts.
It not only considers the probability of occurrence of words and text chunks of different lengths in the reference text dataset, but it also penalizes for duplicate text and inappropriate sentence lengths.
\par

\item\textbf{Fréchet Inception Distance (FID)}. FID \cite{fid} is a commonly used evaluation metric in the field of image generation. 
However, it cannot give a reasonable evaluation of the overfitting situation where the generated results are too similar to the training dataset.\par
\item \textbf{Contrastive Language-Image Pre-Training (CLIP) Score}. CLIP score is an evaluation metric in the field of cross-modal generation or retrieval, which evaluates the matching situation of text and images. It extracts the features of text and image by the multi-modal model CLIP \cite{clip} and then calculates the cosine similarity of the features. The higher the feature similarity, the better the match between image and text. 

\item\textbf{Age of Information (AoI).} AoI is a metric that assesses the amount of time between the generation or capture of information and its reception or utilization. It demonstrates the timeliness of the information, or its vitality. 
A low AoI indicates that the information is extremely current, whereas a high AoI indicates that the information is relatively outdated. It is frequently employed in situations where high timeliness of data collection is required, like sensor networks. 

\item\textbf{Generalization}.  
The loss of the DL model on the training set decreases as the number of parameters increases, but decreases and then increases on the test set, which means that the model moves from underfitting to overfitting. 
Consequently, for different DNNs, it is essential to evaluate their accuracy and generalization in a targeted manner and make a trade-off between them. For instance, some DNNs exhibit a second decrease in the loss on the test set after over-parameterization \cite{tradeoff}.\par
\end{itemize}
 \emph{b) Metrics Related to Communication Transmission:} When applying LLMs to the communication field, it is necessary to focus on their characteristics related to communication, so as to design new metrics about the transmission performance, which will make LLMs more adapted to the communication environment and work better.\par
\begin{itemize}
\item\textbf{Delay}. The amount of data to be transmitted can be reduced by utilizing LLMs to extract semantic information or concise prompts from different forms of data for communication transmission. At this point, the LLM can be viewed as a lossless compressor on the transmitted data, which implements compact representation of the data using a DL model based on the redundancy of human language. 
Thus, the higher the compression rate in the large model, the lower the decoding delay, but the performance must drop accordingly, corresponding to the fact that increasing the performance and decreasing the compression rate will result in a longer decoding delay.\par
\item\textbf{Distortion}. 
AIGC utilizes AI models to automatically generate various types of text, images, audio, video and other content according to given conditions such as themes, keywords, formats, and styles. The generated content by AIGC may lack personalization compared to the works of humans.
In the case of LLMs (large-scale LMs), given that LMs are fundamentally probabilistic in nature, employing probabilistic metrics to assess model performance, such as the Wasserstein distance for measuring distortion, may offer a more suitable criterion.

\item\textbf{Bandwidth Utilization.} Focusing on the network bandwidth occupied by AIGC-generated content ensures that network resources are effectively utilized. On the one hand, when bandwidth resources are underutilized, a dynamic bandwidth allocation mechanism that adapts to changing network conditions and traffic demands can be implemented. 
Moreover, a large quantity of AIGC data may exceed the communication bandwidth. 
\end{itemize}
\subsection{Lessons Learned}
Building a mobile AIGC network for distributed AIGC devices can effectively support the demands of large-scale content training, generation, and distribution. This network will facilitate seamless connectivity among devices, enabling efficient collaboration and resource sharing. Moreover, the low latency and high bandwidth achieved through effective network resource management present significant opportunities to enhance the real-time responsiveness of LLMs. By harnessing these benefits, the performance of AIGC applications can be improved, ensuring they meet diverse user demands in dynamic contexts.

\section{Applications}\label{s6}
In this section, we present several applications of LLMs and AIGC technologies in communication networks, including intelligent communication networks, human machine dialogue interaction, smart home, and case studies. 
\subsection{Intelligent Communication Networks}
The introduction of AIGC and LLMs in the field of communications holds new promise for realizing a more intelligent communication network. They are playing a key role in semantic communication, network design,resource management, and pushing the frontiers of communication technologies.
As the most powerful NLP technologies currently, LLMs greatly facilitate semantic conveyance in communication. This enables the system to accurately interpret and generate semantically meaningful communication content. As a result, the communicating parties are able to understand each other's intentions more clearly during the information exchange process, thus reducing misunderstandings and communication barriers. Moreover, the generative models show excellent ability in adaptive processing of signal noise.
In \cite{letafati2023denoising}, the authors point out that there is a close relationship between the underlying mechanisms of diffusion models and the denoising, decoding, and reconstruction of information signals by the receiver of a communication system in response to noisy and distorted received signals. 

Considering the notable achievements of diffusion models in the realm of decision optimization, AIGC provides an idea for liberating the huge operation and maintenance costs in network design.
In \cite{huang2023networkdesign}, a network design paradigm based on diffusion models and reinforcement learning is proposed, which can adapt to incomplete network system models and unknown constraints to intelligently generate network solutions quickly based on user intent. The efficacy of this scheme in guiding transmit power allocation in digital-twin-based access networks has been confirmed through simulations. 
Moreover, a DGM-based wireless network management framework is proposed in \cite{liu2023management}. The authors analyze that the excellent generative and feature representation capabilities of GAI models help to model the network state, learn potential features, generate network management policies, and can migrate the learned knowledge to cope with environmental changes. Thus the framework is effective for representative use cases in wireless network management such as network routing, resource allocation, and network economics.

The powerful analysis and generation capabilities of LLMs and AIGC enable them to be applied to solve many scenarios in network security. In particular, LLMs can learn normal patterns of network traffic, user behavior, and system logs. By comparing new data with these learned patterns, they can detect anomalies that may indicate potential threats. In \cite{ferrag2023revolutionizing}, the author proposed SecurityBERT, which utilizes the bidirectional encoder representations from Transformers (BERT) model to detect network threats in IoT networks. During the training process, it employs a novel privacy-preserving encoding technique to effectively represent network traffic data in a structured format. In addition, LLMs can analyze software code, system configurations, and network architectures to identify potential security vulnerabilities. By leveraging various prompt technologies, LLMs would achieve software vulnerability detection \cite{wu2023characterizing}.
\subsection{Human Machine Dialogue Interaction}
Currently, human-machine dialogue products and applications that use text or speech interaction are widely studied due to benefits such as efficiency and accuracy \cite{app1, app2, app3}.
LLMs, with their powerful ability of language understanding and natural fluent dialogue, are expected to improve the interaction experience and promote the development and application of human-computer dialogue.
The correct execution of tasks needs dialogue systems to be able to reason correctly about the intentions of users.
Errors in determining user intent can result in serious accidents in security-related settings. 
Moreover, some tests have shown that advanced LLMs (e.g., GPT-4) have a high level of a theory of mind in understanding user emotions and intentions \cite{zhang2024prompt}, which suggests that LLMs can help dialogue systems achieve an understanding of human minds. 

Broadly, human-machine dialogue can be divided into two types: task-based dialogue and non-task-based dialogue \cite{chen2017survey}. 
In task-oriented dialogue, the system primarily focuses on carrying out tasks, requiring it to comprehend, inquire, and further clarify users' needs to ensure they are adequately met. 
Non-task-based dialogue focuses on conversing with human \cite{chen2017survey} and provides users with rich information and fluent responses. 
Thanks to their advanced language processing capabilities and rich knowledge base, LLMs can promote the development of non-task-based dialogue systems. 
In the field of programming, there exists a conversational programming assistant supported by LLMs that can improve the productivity of software engineers \cite{ross2023programmer}. 
This LLMs-based assistant can answer common programming questions, generate context-sensitive code, and even ask follow-up questions about the user's session and code context. 

\vspace{-0.2cm}
\subsection{Smart Home}
\begin{figure}[!t]\vspace{-0.3cm}
	\centering
	\includegraphics[width=0.7\linewidth]{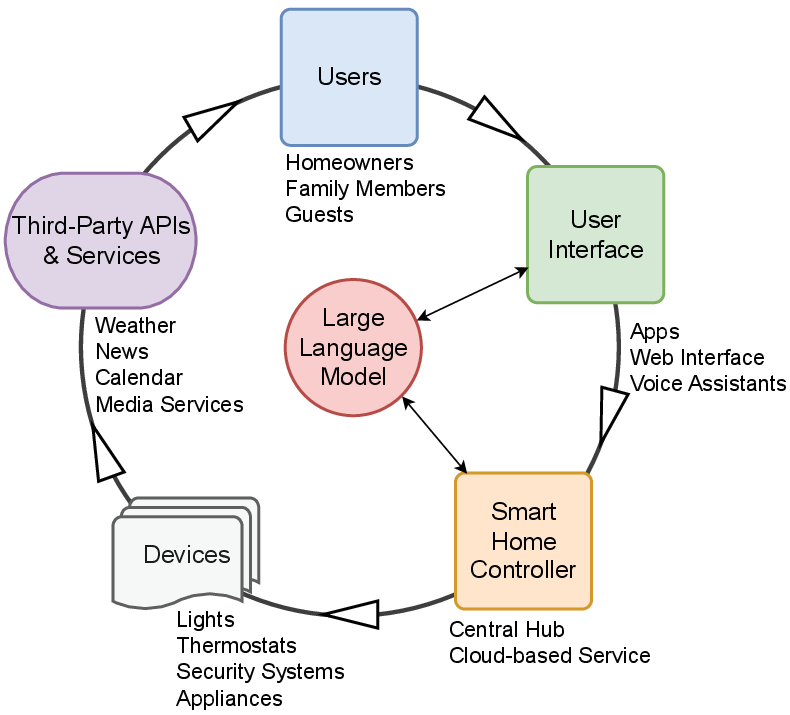}
	\vspace{-0.3cm}\caption{A cutting-edge structure that fuses smart home with LLMs.}
	\label{Application_1}
\end{figure}

Smart homes \cite{robles2010applications, hasan2018smart, wu2022characterizing} utilize Internet of things (IoT) technologies to interconnect home amenities, appliances, and services for streamlined daily operations. Improvements to the communication network will significantly enhance the smart home system's real-time response, stable connectivity and secure transmission. Current smart home systems consist of multiple subsystems, each focusing on specific tasks or devices, such as voice assistants, smart refrigerators, smart vacuums, and security door locks. However, the lack of central autonomous control and command (C\&C) systems hinders cooperation among subsystems \cite{hammi2022survey}. Integrating the context-aware conversational capabilities of LLMs with home automation enhances distributed single entities by creating an organic ecosystem of devices, which is a prototype of a small mobile AIGC network.
Fig.~\ref{Application_1} depicts an innovative architecture that combines a smart home system with a LLM. In this architecture, the user interacts with the system via a user interface. The global context-aware controller transmits the user's commands to the respective smart home devices. By leveraging the context-aware dialog capabilities of LLMs, the user experience is optimized. Additionally, it can integrate with third-party APIs and services to broaden its functionality and offer additional information and services.

From the perspective of users, such as homeowners, family members and guests, LLMs enable more natural voice control and personalized modeling. First, due to the powerful context-aware dialogue capability of LLMs, it can well parse the implied intent of user voice input, improving the naturalness and intelligence of the interaction and creating a simpler and more comfortable smart home experience for users. 
Existing speech interaction designs are limited to fixed predefined commands, which raising the bar for users \cite{noura2020natural}. 
Second, LLMs can analyze large amounts of time-series data from smart hardware and sensors to reveal user habits and behavior patterns. It can also analyze the user's scenario by identifying information such as location, time and device status to provide different services and improve user satisfaction and experience towards smart home services.

From the perspective of ecological environment, combining a LLM with a smart home system enables intelligent operation, management, and improved security control \cite{noura2020natural}. By analyzing data on household water, electricity, and gas usage, LLMs can provide accurate solutions and recommendations for energy saving to improve overall energy efficiency \cite{rajesh2022novel}. Additionally, LLMs can analyze data from smart home devices and appliances proactively to identify potential problems early. LLMs can also learn common threat patterns to enhance security control in smart home systems \cite{taiwo2022enhanced}, including automatic identification of intruders or potential security risks.
\begin{figure*}[!t]\vspace{-0.8cm}
\vspace{0cm}
	\centering
	\includegraphics[width=0.75\linewidth]{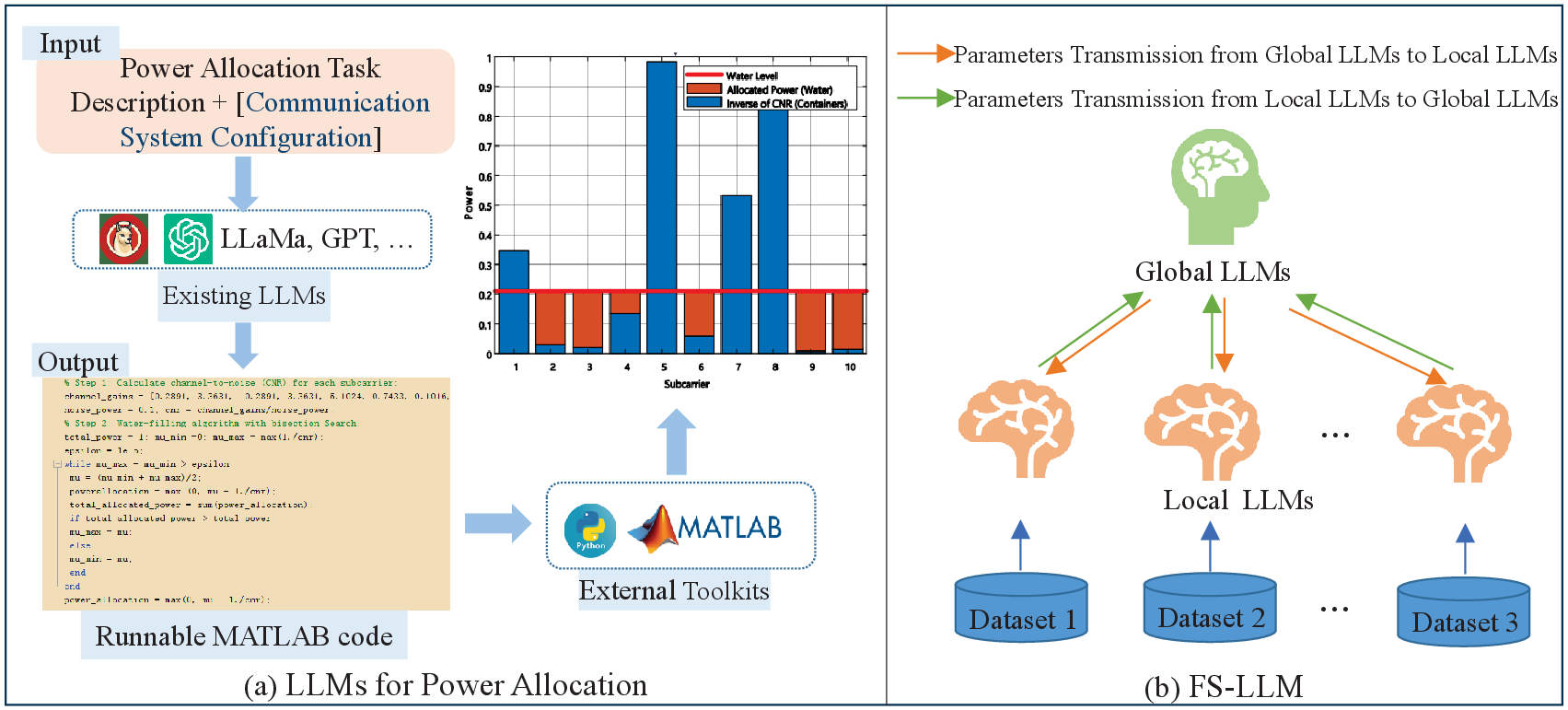}
	\vspace{-0.2cm}\caption{The case studies of communication-specific LLMs and communication network deployment for LLMs.}
	\label{Case}
\end{figure*}

\subsection{Case Studies}
To demonstrate the feasibility of
the application of LLMs in the field of communication, we conduct the following case
analysis. 
Specifically, 
\begin{itemize}
    \item \textbf{Communication-Specific LLMs:} Currently, many researchers are working on single LLM-based methods to deal with multiple tasks in the field of communication. In \cite{shao2024wirelessllm}, the authors employ existing LLMs such as GPT and LLaMa to realize different types of tasks in wireless communication including power allocation, spectrum awareness, and network protocol understanding through prompt engineering and retrieval enhancement generation techniques. 
    Fig.~\ref{Case} (a) shows a case of how LLMs solve the power allocation problem. In particular, for a given communication scenario, relevant model parameters such as the number of subcarriers, the channel gain corresponding to each subcarrier, and the noise power need to be set and input into existing LLMs along with the description of the power allocation task. Based on the input information, LLMs divide the complex power allocation task into several simple sub-tasks through step-by-step analysis and then generate Matlab code. The final calculated solution is obtained by inputting the code into the Matlab interpreter. 
    However, existing LLMs learn relatively limited communication-related knowledge from the existing datasets, resulting in low accuracy when solving tasks in communication scenarios. To address this issue, the authors in \cite{zou2024telecomgpt} propose TelecomGPT, which trains LLMs in three steps (i.e., continual pre-training, instruct tuning, and alignment tuning) and establishes corresponding datasets for each training step.
    Moreover, the evaluation benchmarks are proposed to measure the performance of LLMs in the field of communication.
    \item \textbf{Communication Network Deployment for LLMs:} LLMs may need to be fine-tuned on different users' data for solving different tasks. However, due to privacy issues, users' local data cannot be directly shared, and there are many challenges in fine-tuning LLMs within FL framework, including high computational cost and the lack of standardized benchmark for performance evaluation. 
    Consequently, the authors in \cite{kuang2024federatedscope} propose an FederatedScope-LLM (FS-LLM) framework that integrates multiple datasets, diverse fine-tuning algorithms, and corresponding evaluation metrics. Fig.~\ref{Case} (a) shows the FS-LLM framework, which consumes only a small amount of computing and bandwidth resources and is an excellent approach for deploying LLMs in devices with constrained computing resources. As FL conducts joint training on the data of each user through a communication network, the performance of LLMs are significantly improved compared to an LLM that solely utilizes local data.  
\end{itemize}

\section{Challenges and Future Directions}\label{s7}
In this section, we discuss the challenges and future directions from two perspectives: foundation models of communication and AIGC and LLMs on end devices.
\subsection{Foundation Models of Communication}
\subsubsection{General Intelligence For Communication}
 Traditional DL methods applied to the communication field have favourable performance in task-oriented applications, but it is difficult to achieve artificial general intelligence (AGI). In a classic communication process, it may include tasks such as source-channel coding, data processing, and privacy protection. It is usually time-consuming to find a certain matching neural network for each task among numerous neural network structures. A general AGI system can greatly improve the efficiency and reduce the cost of model development. Building a foundation model of communication is the basis for realizing the general intelligence for communication. Meanwhile, prompt engineering can establish the internal relationship between specific tasks and the knowledge learned by the pre-trained foundation model. By modifying the prompts, the model can quickly adapt to new scenarios and new requirements.
\subsubsection{Model Hallucination Phenomenon}
Model hallucinations \cite{rR12} can be categorized into factuality and faithfulness hallucinations. Factuality hallucinations are those in which the model's output is inconsistent with real-world information, including common-sense errors and fictionalized content, while faithfulness hallucinations involve logical errors, such as deviation from user instructions or improper reasoning. The causes of these hallucinations lie primarily in data and modeling deficiencies.\par
Data deficiencies include data bias, outdated information, insufficient data in specialized domains, and underutilization of parametric knowledge and overreliance on knowledge co-occurrence thus limiting reasoning. Therefore, it is particularly important to improve data quality and the ability of models in learning and memorizing factual knowledge. For example, knowledge graphs are a structured way of representing knowledge and can provide accurate knowledge and information for the model. Combining knowledge graphs with LLMs enables the model to refer to the entities, relationships, and attributes in knowledge graphs when generating text, thereby improving the accuracy and reliability of generated text and reducing hallucinations.\par
In terms of model deficiencies, the Transformer-based architecture uses unidirectional attention, which may lead to underrepresentation and attentional dilution for long sequences. Decoding mechanisms in coding and decoding architectures often employ randomization and local contextual attention strategies, which limit the model's veracity and expressiveness to some extent. During training, the autoregressive model generates erroneous outputs that serve as inputs for subsequent training, thus creating a snowball effect. In addition, RLHF often caters to human preferences, but this can lead to the output of inaccurate information. To address the above issues, the model architecture can be improved or the model's understanding of factual associations can be enhanced by fact-augmented training strategies. Retrieval-augmented generation (RAG) provides a way to utilize retrieval results to aid in the generation of answers by utilizing more external knowledge in the decoding phase, thereby reducing the phenomenon of illusions.
\subsubsection{Theory of LLMs}
Currently, LLMs are still an engineering-based technology and lack the theoretical basis to expose their principles. Therefore, it is still challenging to construct and optimize large models effectively. Advancing LLMs from engineering-based to scientific theory-based will help us tremendously in using and constructing LLMs for communication networks.
\begin{itemize}
    \item Emergence Ability:
        The emergence ability of LLMs means that as the number of model parameters grows substantially, it demonstrates abilities that are absent in smaller models. Investigating the principles underlying the emergence phenomenon is a critical area of focus. This can be artificially induced when confronted with limited parameters and insufficient dataset.
    \item Chain-of-Thought:
        For a complex problem, it is difficult to generate appropriate results from LLMs if only the problem description is given. However, if the thought process of a question is used as a prompt, relatively high-quality generated content can be produced. Chain-of-thought prompting can stimulate the reasoning power of LLMs to solve some of the more complex problems, which makes LLMs more explainable and more credible. 
    \item Evaluation and Optimization:
        The quality of the generated contents largely depends on the user's personal use experience. In the evaluation of the GAI models, more user’s feedback needs to be introduced. It is a good direction to find a suitable way to add the user's personal evaluation to the training of the model according to a certain percentage. By using a evaluation scheme with user's proposal, the realism of the generated data will be increased, while greatly improving the user experience.
\end{itemize}
\subsection{AIGC and LLMs on End Devices}
\subsubsection{Insufficient of Computility and Storage Capacity}
End devices lack computing power and storage capacity, which makes it difficult to deploy LLMs and AIGC applications. Many tasks in the communication domain require high real-time performance, and thus will result in high demands on the inference speed of the model.
GAI models generally possess complicated structures, especially LLMs with billions of parameters, which introduce challenges in their training process and inference timing when combined with communication technologies. 
Current LLM-related applications use cloud computing to solve the problem of lack of computing power and storage capacity, but this can lead to more problems, such as high latency, high bandwidth, and privacy issues. If the computing steps are properly distributed to the cloud devices, and the edge devices and end devices, these problems will be improved. Split Learning (SL) allocates computing power by decomposing large-scale models, but its training efficiency is low and difficult to deploy. Therefore, the research of SL is an important issue. Besides, multi-agent LLM Network is also a good research direction to improve the quality of end AIGC service through the cooperation of multiple edge devices.
\subsubsection{Communication Costs and Latency}
LLMs require a lot of communication resources in the process of inferring and training when response to edge devices. When a user uses some applications on an end device, it is inevitable that there will be multi-media information interaction. But the available bandwidth in communication is limited, which can cause a lot of network latency and affect the user experience. Since the GAI model has strong content generation capabilities, we can generate the multi-media information to transmit by using brief prompt. However, the content generated by GAI is unstable, and the generated content may not be the data we originally transmitted. How to accurately generate the information we truly need is a very important direction.
\subsubsection{Personal Privacy Protection}
While LLMs and AIGC technologies can be used to optimize potential data security issues in communication systems, big models themselves are subject to privacy breach risks and data. 
On the one hand, data collected in various ways for training LLMs may contain user personal information, leading to privacy leakage. Besides, there are sometimes copyright issues with web page data. On the other hand, AI-generated data may contaminate real data. With the development of LLMs and AIGC, it is more and more convenient to generate content with AI, and the authenticity gradually increases, which is difficult for humans to distinguish. This not only represents the advancement of AI technology, but also brings the crisis of the proliferation of false data. Moreover, the selection of data and algorithms can bring bias in the generated content, resulting in serious social problems such as public opinion guidance or the proliferation of negative information. 
\section{Conclusions}\label{s8}
In this paper, we first introduce the background of LLMs and other AIGC technologies, and provide an overview of traditional AI-empowered communication technologies. Then, we review the technologies related to the integration of LLMs and AIGC with communication networks from two perspectives that mutually reinforce each other. Specifically, we reveale the contribution of LLMs and AIGC in terms of data augmentation, communication scenario and network configuration generation, semantic communication, and security and privacy protection. Besides, the improvement of distributed training, cloud-edge-terminal collaboration, and network resource control to AIGC services and the performance of LLMs were discussed.
We also summarize the applications, including intelligent communication networks, human machine dialogue interaction, smart home, and smart healthcare. At last, we discuss the challenges and future directions. We expect that the insights provided by this survey will shed light for other researchers on the current status and future directions of the integration of communication networks with LLMS and AIGC technologies, and motivate more explorations and investigations in this intersecting research field.
\bibliographystyle{IEEEtran}
\bibliography{manuscript}
\end{document}